\documentclass[reprint,amsmath,nofootinbib,showkeys,fleqn,amssymb,aps,prd,superscriptaddress]{revtex4-2}

\usepackage{graphicx}
\usepackage{dcolumn}
\usepackage{bm}
\usepackage{xcolor}
\usepackage{orcidlink}

\newcommand\farcs{\hbox{$.\!\!^{\prime\prime}$}} 

\begin{document}

\preprint{APS/123-QED}

\title{Improved dark matter measurements with flexible modeling of \\ resolved strongly-lensed quasar narrow-line emission}

\author{Maria F. Perez Mendoza, \orcidlink{0009-0005-5078-117X}}
\author{Anna M. Nierenberg, \orcidlink{0000-0001-6809-2536}}
\email{anierenberg@ucmerced.edu}
\affiliation{%
 University of California, Merced, Merced, CA, USA
}

\author{Vardha~N.~Bennert\,\orcidlink{0000-0003-2064-0518}}
\affiliation{Physics Department, California Polytechnic State
  University, San Luis Obispo, CA 93407, USA}

\date{\today}

\begin{abstract} 

The relative brightnesses of strongly lensed quasar images, called flux ratios, respond to perturbations from low-mass dark matter halos, enabling tests of dark matter models. The quasar narrow-line region (NLR) is ideal for flux-ratio studies: large enough to be insensitive to stellar microlensing, yet compact enough to remain sensitive to dark matter halo substructure. While nuclear emission dominates NLR flux, many quasars show low surface brightness extended emission spanning kiloparsec scales that could bias measurements. To test this potential bias, we generated mock Keck OSIRIS AO observations of seven $z<1$, $L_\mathrm{bol}\sim10^{46}$\,erg\,s$^{-1}$ quasars characteristic of sources. Only one system shows detectable extended emission after lensing. We introduce a new pipeline for simultaneously fitting point sources (nuclear) + Sérsic elliptical profiles (extended [O\,III]). We show that we recover the true flux-ratios to $<5\%$ even when the extended emission is boosted to 100 times its original flux. We also demonstrate that visual inspection of lenses reliably determines whether to use point-source-only or include extended emission modeling in the pipeline; both achieve $<5\%$ accuracy--- which is below the typical spectral fitting precision. The new pipeline and fitting procedure ensures reliable flux-ratio measurements can be made of narrow-line flux ratios for the thousands of lenses which will be discovered by Euclid, Rubin and Roman Space Telescopes.

\end{abstract}

\maketitle



\section{\label{sec:background} Background}

Dark matter (DM) constitutes the majority of matter in the Universe \cite{Aghanim2020}, yet its fundamental nature remains elusive. Numerous particle candidates have been proposed, including cold dark matter (CDM), warm dark matter (WDM), sterile neutrinos, and axions \cite[see e.g.][ for a summary of candidates]{review_of_part_22}.
Many of these candidates predict distinct dark matter halo mass functions and internal halo density profiles, particularly at sub-galactic mass scales below $10^8\,M_{\odot}$ \cite{Lovell2014, Weinberg2015}. 

Strong gravitational lensing offers a direct and purely gravitational measure of the underlying mass distribution, insensitive to the baryonic content or emissivity of the lens; for a full review of gravitational lensing as a probe for dark matter, see \citep{Vegetti2023}. In particular, strong gravitational lensing is highly sensitive to perturbations induced by low-mass dark matter halos along the line of sight, providing a means to constrain the small-scale structure of dark matter. The flux-ratio anomaly technique leverages this sensitivity by comparing the relative fluxes of multiply imaged quasars \cite{Mao1998, Dalal2002}. 
In a smooth lens potential, the image positions and relative magnifications are determined by the first and second derivatives of the potential, respectively. Therefore, deviations in observed flux ratios from smooth model predictions serve as a robust diagnostic of substructure and a test of dark matter models. This method has been used to place some of the tightest constraints to date on a variety of dark matter models including Warm Dark Matter \cite{Keeley2024, Keeley2025, Gilman2025}, self-interacting dark matter \cite{gilman_constraining_2023} and axion dark matter \cite{laroche_quantum_2022}.

The narrow-line region (NLR) of quasars provides an ideal emission source for this technique, as it is present in nearly all quasars and extends over tens to hundreds of parsecs. This scale is large enough to be insensitive to stellar microlensing, yet compact enough to remain sensitive to perturbations from low-mass dark matter halos \citep{Moustakas2003}

Previous studies employing the Hubble Space Telescope (HST) grism \cite{Nierenberg2017}, and the Keck OH-Suppressing Infrared Integral Field Spectrograph (OSIRIS) \cite{Nierenberg2014}, 
as well as direct low-redshift resolved imaging of the narrow-line structure \citep{MuellerSanchez2011}
found that the lensed narrow-line emission is dominated by an unresolved nuclear component. \citet{MuellerSanchez2011} measured typical full-width at half-maxima (FWHM) of $\sim50-100$ pc for the nuclear component, while the analyses of lensed sources placed an upper limit of approximately $\sim100$\,pc on the NLR size based on HST data \citep{Nierenberg2017}, and marginally resolved an emission component size of $\sim60$\,pc \citep{Nierenberg2019} with OSIRIS. 

However, high-resolution observations of luminous low-redshift quasars with $z <1$ indicate that the NLR structure can be complex, consisting of a compact nuclear component ($\sim100$\,pc) and a low-surface-brightness extended narrow-line region (ENLR) spanning kpc scales, often showing pronounced clumpiness and asymmetry \citep{Bennert2002, Bennert2006, Humphrey2010, Husemann2014, Husemann2019, Chen2019}. One study measured the typical ENLR size to be $\sim10$\,kpc for a population of quasars with median nuclear [O\,III] luminosity $\log(L_{\mathrm{[O\,III]}}/[\text{erg\,s}^{-1}]) = 42.7 \pm 0.15$ \cite{Husemann2012}. While the central NLR is ionized by the active galactic nucleus (AGN), the extended regions may be partially ionized by star-forming activity in the host galaxy outskirts \citep{Bennert2006}, yet in some cases they remain fully affected by AGN activity \citep{Schmitt2003, Greene2011, VillarMartin2011}.

In this work, we assess how such extended emission features might affect flux-ratio measurements used for DM constraints. Using spatially resolved spectroscopic imaging of $z <1 $ quasars matched in luminosity to our lens sample ($L_{\rm{bol}} \sim10^{46} \text{erg s}^{-1}$), we generate a sample of mock lenses. We also introduce a new method for extracting the amplitude of unresolved nuclear emission---our observable used in DM inference---in the presence of ENLR flux. We then simulate strong lensing of these real sources to quantify the accuracy and precision with which the point-source flux component can be recovered. 

This paper is organized as follows. Section~\ref{subsec:sourceSelection} outlines the sources we chose and important information about them. Section~\ref{sec:genMockLenses} describes the simulation pipeline. Section~\ref{sec:fittingObs} presents the new analysis method developed for fitting gravitational lenses that exhibit extended emission. Section~\ref{sec:results} discusses the implications for flux-ratio analysis and DM constraints.

\section{\label{subsec:sourceSelection} Source Selection}

Our objective is to select data for sources with bolometric luminosities $L_{\rm{bol}}$ comparable to those of known galaxy-scale quadruply lensed quasar systems, which typically have $L_{\rm{bol}} \sim10^{46}$ erg s$^{-1}$ and above \cite{Granato1996}. Furthermore, we require that these quasars have existing spatially resolved spectroscopy covering the $\lambda4959$, $\lambda5007$ [O\,III] doublet, as this is a commonly used emission line in lensing studies using the NLR. After making these selections, we are left with a final sample of seven quasars. Table~\ref{tab:sourceTable} lists key information about the sources, including $f_{\text{[O\,III]}}$, $L_{\text{bol}}$, and $z$.

The data for six of the quasars (3C~273, PG~0026+129, PG~1211+143, PG~1426+015, PG~1617+175, and PG~2130+099) come from a study on AGN black hole masses \cite{Winkel2025} (hereafter Winkel25) under program ID 097.B-0080(A). These observations were conducted with the Multi Unit Spectroscopic Explorer (MUSE; \cite{Bacon2010}) on the Very Large Telescope (VLT). All data were acquired in wide-field mode, providing a field of view of $1'\times1'$ with spatial sampling of 0\farcs2 and a spectral range of 4750--9300~\AA\ at a spectral resolution of $R\sim2500$. The targets were observed across multiple nights following consistent observing strategies, with total integration times ranging from 2800~s to 4500~s and employing standard dither-and-offset patterns. All observations were carried out in March, April, and July 2016 under gray-moon and clear-sky conditions, with seeing between 0\farcs6 and 1\farcs0.  

Our final source, HE~1126-0407, was observed with MUSE in wide-field mode (WFM) on 2015 July 11 as part of program 095.B-0015(A), the Close AGN Reference Survey (CARS; \cite{Husemann2022}). The total on-source integration time was 1800~s. Observations were conducted under clear skies with seeing of 0\farcs7. The photometric quality flag indicates good flux calibration throughout the exposure. 

For our analysis, we are particularly interested in the extended [O\,III] emission. For each source, we construct an [O\,III] emission-line map from the full data cube. For sources from the CARS survey, we use the pre-computed [O\,III] maps directly. For the remaining objects, we generate emission maps following a similar procedure to that adopted in the CARS analysis, namely by integrating PSF-subtracted datacubes over the wavelength range encompassing the [O\,III]~$\lambda5007$ line. In all cases, the resulting maps are scaled by a factor of $4/3$ to account for the combined contribution of the [O\,III]~$\lambda5007$ and [O\,III]~$\lambda4959$ transitions. To mitigate residual PSF artifacts near the center, negative pixels are masked and replaced with the mean value of surrounding regions. 

\begin{figure*}
    
\includegraphics[width=1\textwidth]{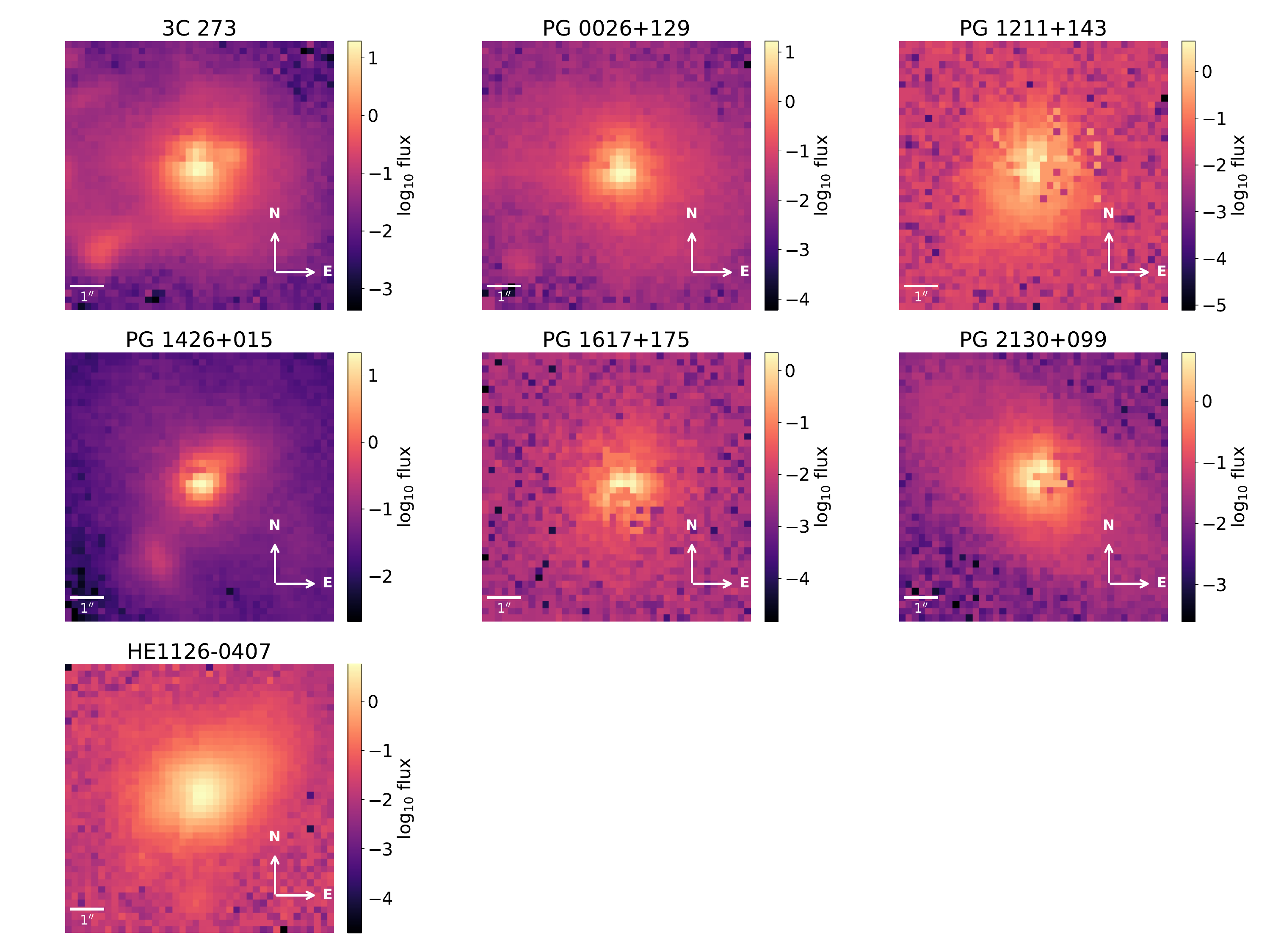}
\caption{\label{fig:oiiiMaps} Visualization of the unlensed [O\,III] emission reveals sources with complex morphological features, including elongation, skewness, and clumpiness. The detailed properties of these sources are summarized in Table \ref{tab:sourceTable}. }
\end{figure*}

The resulting point-source-subtracted [O\,III] maps are shown in Fig.~\ref{fig:oiiiMaps}, revealing complex morphologies including elongated structures and asymmetries. This cleaning is essential to accurately characterize the extended [O\,III] emission without contamination from the bright quasar point source which appears as a point spread function (PSF).

\begin{table*}
\caption{\label{tab:sourceTable}Table with the source properties coming from Winkel25 \cite{Winkel2025} and CARS \cite{Husemann2022}.}
\begin{ruledtabular}
\begin{tabular}{ccccc}
 AGN Name & Survey &$f_{\text{[O\,III]}}$ & $L_{\text{bol}}$ & \textit{z} \\
& & [$10^{-16}$ erg s$^{-1}$ cm$^{-2}]$ & [$10^{45}$ erg s$^{-1}$ ] & \\

\hline
3C~273 & Winkel25 & $ 711 \pm 9$ & $ 9600 \pm 600$ & $ 0.160$ \\ 
HE~1126-0407 & CARS & $668 \pm 2$ & $ 230 \pm 20$ & $0.060$ \\
PG~0026+129 & Winkel25 & $540 \pm 10 $ & $ 710 \pm 60$ & $0.140$ \\
PG~1211+143 & Winkel25 & $834 \pm 1$ & $ 400 \pm 10$ & $0.080$ \\
PG~1426+015 & Winkel25& $428 \pm 1$ & $ 540 \pm 10$ & $ 0.086$ \\
PG~1617+175 & Winkel25& $102 \pm 2$ & $ 600 \pm 15$ & $0.110$ \\
PG~2130+099 & Winkel25 & $563\pm 1$ & $251 \pm 2$ & $0.063$ \\

\end{tabular}
\end{ruledtabular}
\end{table*}

\section{Generating Mock Lenses}\label{sec:genMockLenses}

In this section, we describe the procedure used to generate mock OSIRIS observations in the H broad band (1.47–1.80~$\mu$m; \citealt{osiris}), representative of near-infrared, adaptive-optics–assisted data commonly employed in strong-lensing studies.
We construct a two-component simulation consisting of a resolved plus an unresolved component.  We treat the case in which the extended [O\,III] emission was obtained by lensing the [O\,III] maps from our data set, while the unresolved quasar component was modeled as a point source. Because of the distinct spatial characteristics of these components, we treat the extended and point-like emission with different modeling strategies, and we describe the detailed steps for each component in the subsequent subsections. All lensing simulations are performed with the open-source software package \texttt{lenstronomy}~\cite{Birrer2018,Birrer2021}.

\subsection{Macromodel}\label{subsec:macromodel}

We adopt a standard macromodel for the lens system, consisting of a Singular Isothermal Ellipsoid (SIE) with external shear. This choice is motivated by the fact that SIE+shear models have been shown to provide a good description of the combined stellar and dark-matter mass profiles of elliptical galaxies \cite{Treu2010}. We adopt typical values measured for quadruply imaged quasars \cite{Schmidt2022}  with an Einstein radius of 
\(\theta_{\mathrm{E}} = 0\farcs7\), position angle $\phi = -11.6 ^\circ$ (N of E), axis ratio $q = 0.86$, and external shear angle $\phi_{\text{ext}}= -45.0 ^\circ$,  (N of E) and external shear strength $\gamma_{\text{ext}} = 0.05$. 

We also investigate the impact of different lensing configurations by varying the position of the background source in the source plane relative to the lens mass centroid. We consider three representative configurations: for the cross configuration, the source is placed at \((0.0, 0.0)\); for the cusp configuration, at \((0\farcs042, 0\farcs0125)\); and for the fold configuration, at \((0\farcs02, 0\farcs02)\).

\subsection{Extended Narrow-Line Region Lensing \label{subsec:extendedNLRLensing}}

To simulate strong lensing of ENLR, we redshift the original $z<1$ images to $z=2.26$, where [O\,III] $\lambda 5007$ emission is detectable in OSIRIS $H$-band. The redshifted flux density follows \cite{Hogg1999}:

\begin{equation}
S_{\lambda_\mathrm{obs}} = \frac{L_{\lambda_\mathrm{em}}}{4\pi D_L^2 (1+z)},
\label{fluxRelation}
\end{equation}

where $S_{\lambda_\mathrm{obs}}$ is the observed flux density, $L_{\lambda_\mathrm{em}}$ is the emitted luminosity density, and $D_L$ is the luminosity distance. The flux ratio between redshifts $z_1$ and $z_2$ is then

\begin{equation}
\frac{S_{\lambda_\mathrm{obs},1}}{S_{\lambda_\mathrm{obs},2}} = \frac{(1+z_2)}{(1+z_1)} \frac{D_{L,2}^2}{D_{L,1}^2}.
\label{fluxConversion}
\end{equation}

We transform image fluxes from the original redshift to $z=2.26$ using Eq.~\ref{fluxConversion}, assuming a flat $\Lambda$CDM cosmology ($H_0=69.6$~km~s$^{-1}$~Mpc$^{-1}$, $\Omega_\mathrm{m}=0.286$, $\Omega_\Lambda=0.714$).

The redshifted [O\,III] emission is lensed using the \texttt{INTERPOL} source-light model in \texttt{lenstronomy}. This approach lenses the source as a finely sampled surface-brightness distribution, preserving the complex morphology of the real sources. The result is a lensed image of the extended [O\,III] emission, integrated over the narrow emission-line wavelength range. 

For the final step, we construct a mock OSIRIS datacube by generating a sequence of lensed [O\,III] images with amplitudes scaled according to the [O\,III] doublet spectrum. We generate a simulated spectrum with the double [O\,III] peaks as the only emission and then multiply the lensed image by the corresponding fractional flux at each wavelength. This assumes wavelength-independent spatial distribution of the [O\,III] emission across the spectral region, neglecting potential spatially correlated spectral features (e.g., from winds or rotation). We expect such wavelength-dependent spatial variations to contribute only secondarily relative to the dominant point source flux.

\subsection{Point Sources}\label{subsec:pointSourcesLensing}

The remaining unresolved quasar emission was modeled as originating from a point source. The resulting four image positions and corresponding magnifications are obtained from the macromodel (see \ref{subsec:macromodel}) and convolved with a Gaussian PSF with $\mathrm{FWHM} = 0\farcs1$, consistent with typical AO-corrected OSIRIS data. 

To construct a mock OSIRIS datacube for the point sources, we followed a procedure analogous to that adopted for the extended [O\,III] emission, whereby the two-dimensional lensed image was scaled by the fractional flux in each spectral channel to create the three-dimensional datacube. In this case, the quasar spectrum was extracted from the original, unlensed MUSE data using a circular aperture centered on the peak of the PSF model. The initial spectrum from the MUSE observations is then redshifted, splined, and resampled onto the OSIRIS wavelength channels, and the corresponding lensed images are subsequently scaled by the flux at each wavelength to generate the final three-dimensional point source datacube.

\subsection{Finalizing the Observations}\label{subsec:finalObs}

\begin{figure}
    \includegraphics[width=1\linewidth]{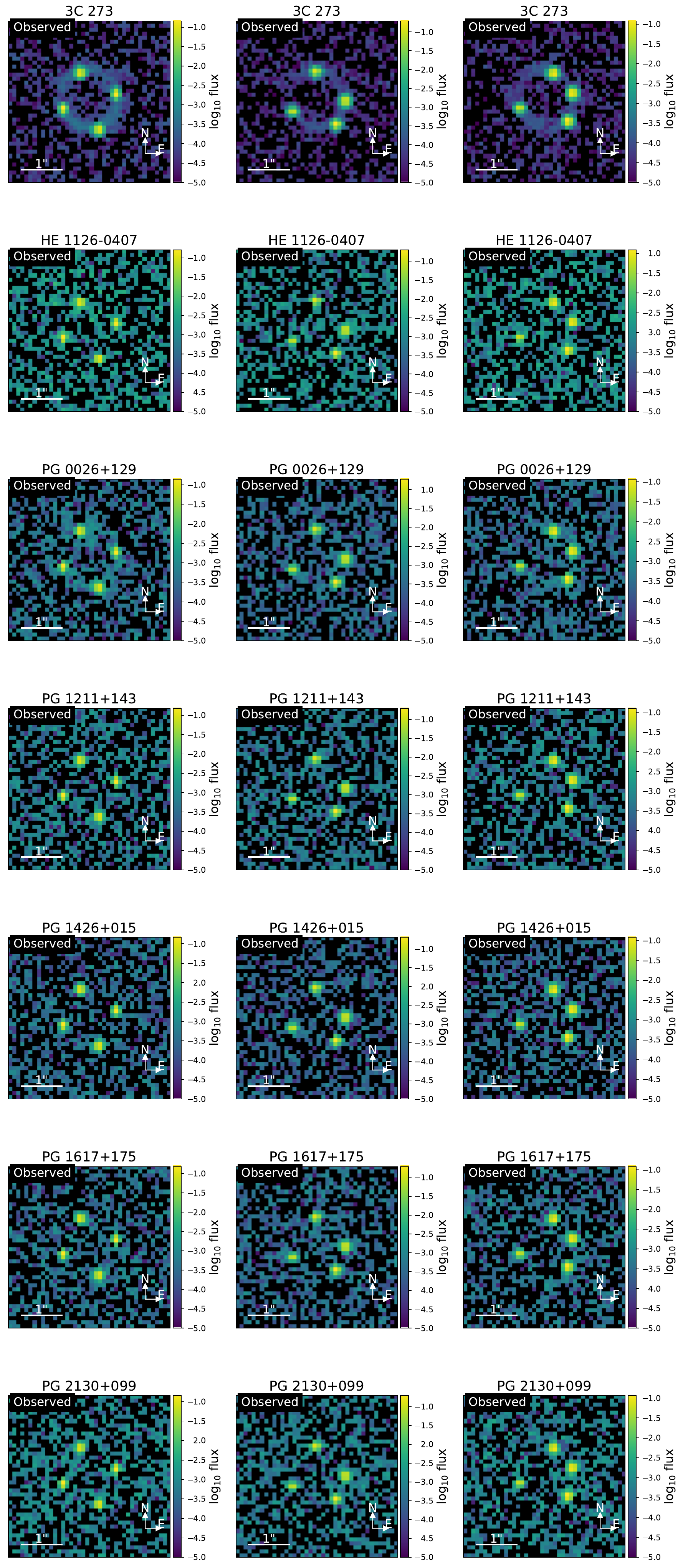}
    \caption{Mock lenses created from the 7 original sources in the cross (left), fold (center), and cusp (right) configurations. Out of these only 3C~273 (top row) has detectable extended [O\,III] emission and thus requires our new two-component (PS + Extended source) fit. The rest can be fit with a point source only model.\label{fig:OGLenses}}
    
\end{figure}

To obtain the final mock OSIRIS observation, we combine the simulated datacubes for the extended [O\,III] emission (\ref{subsec:extendedNLRLensing}) and unresolved quasar point sources (\ref{subsec:pointSourcesLensing}), then add Gaussian noise representative of a typical AO OSIRIS exposure \citep{Nierenberg2014}. The resulting cube corresponds to a single effective integration of 1800~s. In this work, we generate single mock exposures, but real observations comprise multiple dithered exposures, which would improve spatial resolution relative to what is presented here.

For visualization, we construct white-light images by integrating the final cube over a narrow spectral window centered on redshifted [O\,III]~$\lambda5007$, producing narrow-band images of the lenses. These simulated observations (Fig.~\ref{fig:OGLenses}) allow by-eye assessment of extended emission detectability. Among the seven mock lens systems, extended [O\,III] emission is detectable only in 3C~273.

\section{Fitting Procedure}\label{sec:fittingObs} 
\begin{figure*}
    \centering
    \includegraphics[width=1\linewidth]{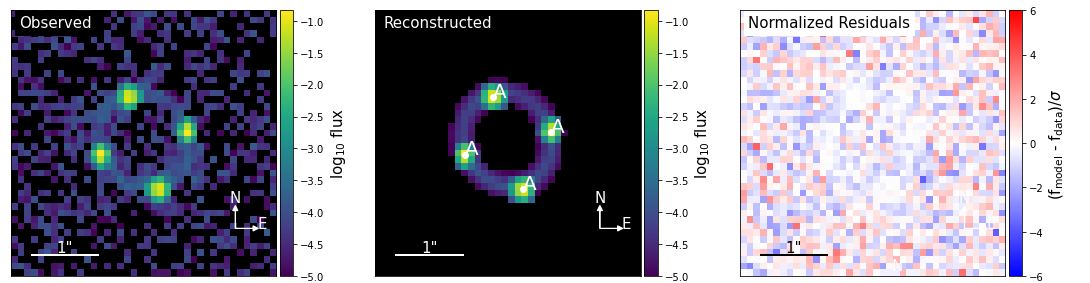}
    \caption{Two-component PS + extended source fit results using our newly developed fitting model for 3C~273, showing one example for the cross configuration (Sec.~\ref{sec:fittingObs}). Left: mock OSIRIS image. Middle: best-fit reconstruction. Right: normalized residuals. Low residual levels indicate an excellent model fit.}
    \label{fig:fitting}
\end{figure*}

In the previous sections, we described how we constructed a sample of mock lenses based on observations of luminous quasars at $z <1$. In this section, we present our method for measuring point-source flux as a function of wavelength in the possible presence of extended [O\,III] emission. The ultimate goal is to extract point-source flux ratios for dark matter studies. The basic concept follows \cite{Nierenberg2014}; here we add model complexity to enable direct modeling of detected extended emission.

We begin by creating white-light images by integrating each datacube over a wavelength interval encompassing the [O\,III] emission. This interval was varied by line width to cover the full width of the redshifted [O\,III]~$\lambda5007$ line (typically $10$--$20$~\AA). These serve as narrow-band images of the lensed system for subsequent lens-model fitting.

We distinguish two regimes based on by-eye inspection, which determines the fitting procedure: systems without significant extended [O\,III] emission use a single-component point-source model following \cite{Nierenberg2014}, assuming perfect PSF knowledge with Gaussian profile as in \ref{subsec:pointSourcesLensing}; systems with clearly visible extended [O\,III] emission use our new PS + Extended Source model.

For the latter---exemplified by 3C~273---the extended [O\,III] emission is described by a S\'ersic elliptical profile \cite{Sersic1967} with surface brightness

\begin{equation}
    I(R) = I_{\rm 0} \exp \left( -b_n \left[ \left( \frac{R}{R_{\rm \text{S\'ersic}}} \right)^{1/n} - 1 \right] \right),
\end{equation}

where $I_{\rm 0}$ is the surface brightness at effective radius $R_{\text{S\'ersic}}$, $n$ is the S\'ersic index, and $b_n \approx 1.999n - 0.327$ ensures half the total light falls within $R_{\rm \text{S\'ersic}}$. The generalized elliptical radius $R$ is

\begin{equation}
    R = \sqrt{q \theta_x^2 + \frac{\theta_y^2}{q}},
\end{equation}

with $q$ as axis ratio and $\theta_x$, $\theta_y$ as angular coordinates. This model (one extra parameter vs.\ Gaussian) balances simplicity and flexibility while capturing essential [O\,III] morphology. Despite complex ENLR morphologies, we find the elliptical S\'ersic profile sufficient for lensed sources.

We use a particle swarm optimizer (PSO) with 20 particles and 1000 iterations per system to explore the parameter space and obtain best-fitting models for both lens and extended emission. PSO is well-suited for this non-convex, multimodal optimization problem, effectively escaping local minima unlike gradient-based methods. An illustrative example of the two-component fit applied to narrow-band images exhibiting extended [O\,III] emission is shown in Fig.~\ref{fig:fitting}. The small residuals demonstrate the robustness of our new PS + Extended Source fitting strategy under representative OSIRIS observing conditions.

\begin{figure*}
    \centering
    \includegraphics[width=1\linewidth]{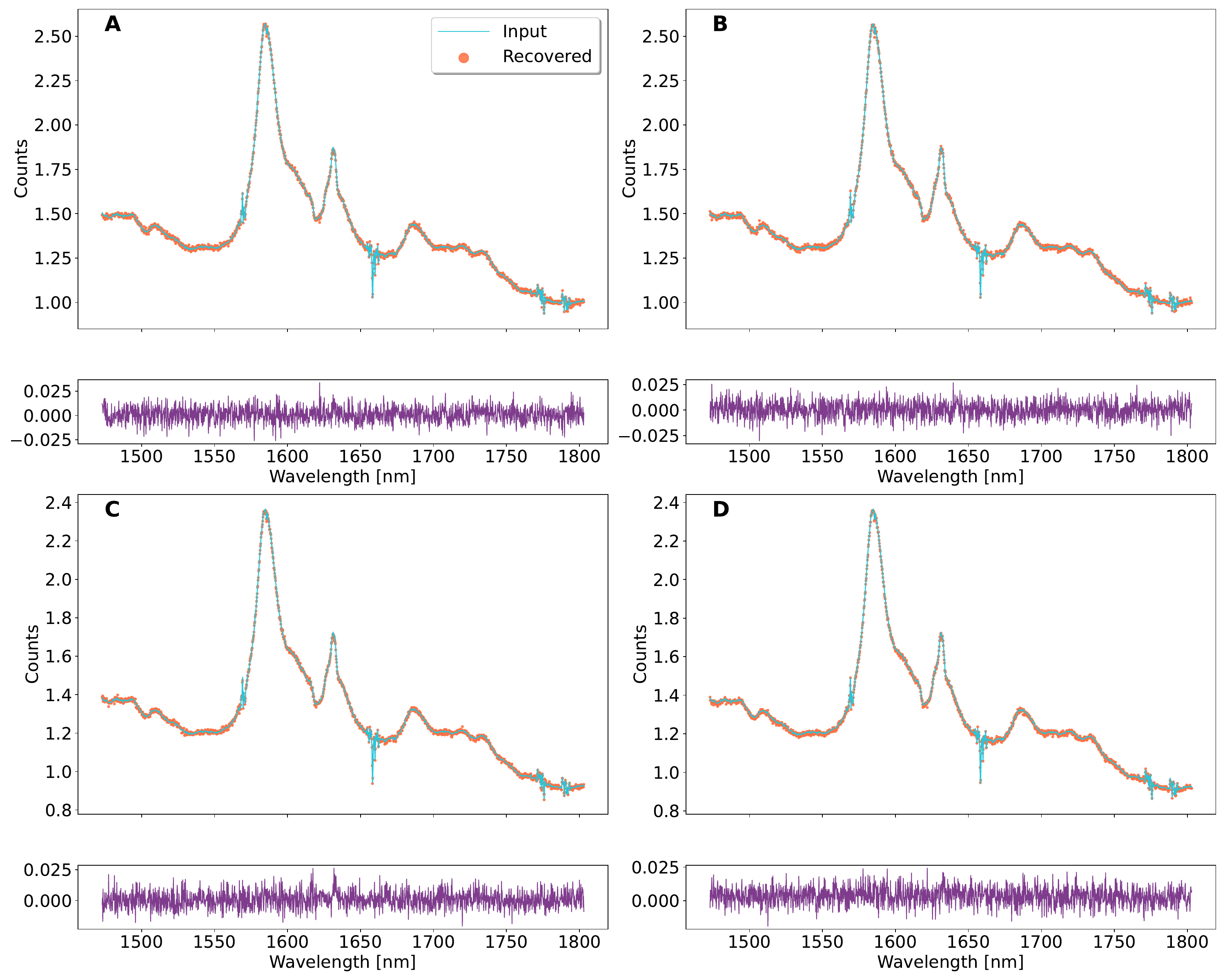}
    \caption{Recovered vs.\ input point-source spectra for 3C~273 images A--D (brightest: A). Blue lines show input spectra (unresolved emission from \ref{subsec:pointSourcesLensing}); orange points show recovered spectra, visually overlapping almost perfectly. Purple regions show residuals (recovered $-$ input). Flux ratios are calculated by integrating flux over [O\,III] $\lambda5007$ wavelength range for each image, then computing ratios relative to brightest image A (B/A, C/A, D/A). This yields ``true'' values (input spectra) and ``recovered'' values (extracted spectra) for comparing pipeline accuracy.}
    \label{fig:inputVsRecSpectrum}
\end{figure*}

To extract spatially resolved spectra for each lensed image, we exploit the wavelength-independent PSF assumption across the narrow [O\,III] range. For fixed model parameters (macromodel + source morphology from spatial fit), we perform weighted linear least-squares optimization per wavelength channel, solving for flux normalizations (amplitudes) of each component (point source and elliptical S\'ersic profile) that best reproduces the observed brightness profile for each lensed image. This yields wavelength-dependent amplitudes, producing a full spectrum for each image.

Fig.~\ref{fig:inputVsRecSpectrum} shows excellent spectral recovery for 3C~273 images A--D. The blue input spectra (unresolved emission from \ref{subsec:pointSourcesLensing}) and the orange recovered spectra overlap almost perfectly, with the signal much higher than the noise across these wavelength ranges. The purple residual subpanels remain flat and near zero, confirming excellent fits for all images A--D even with extended [O\,III] present. This validates our pipeline for flux-ratio anomaly analysis in dark matter studies.
\begin{figure}
    \centering
    \includegraphics[width=1\linewidth]{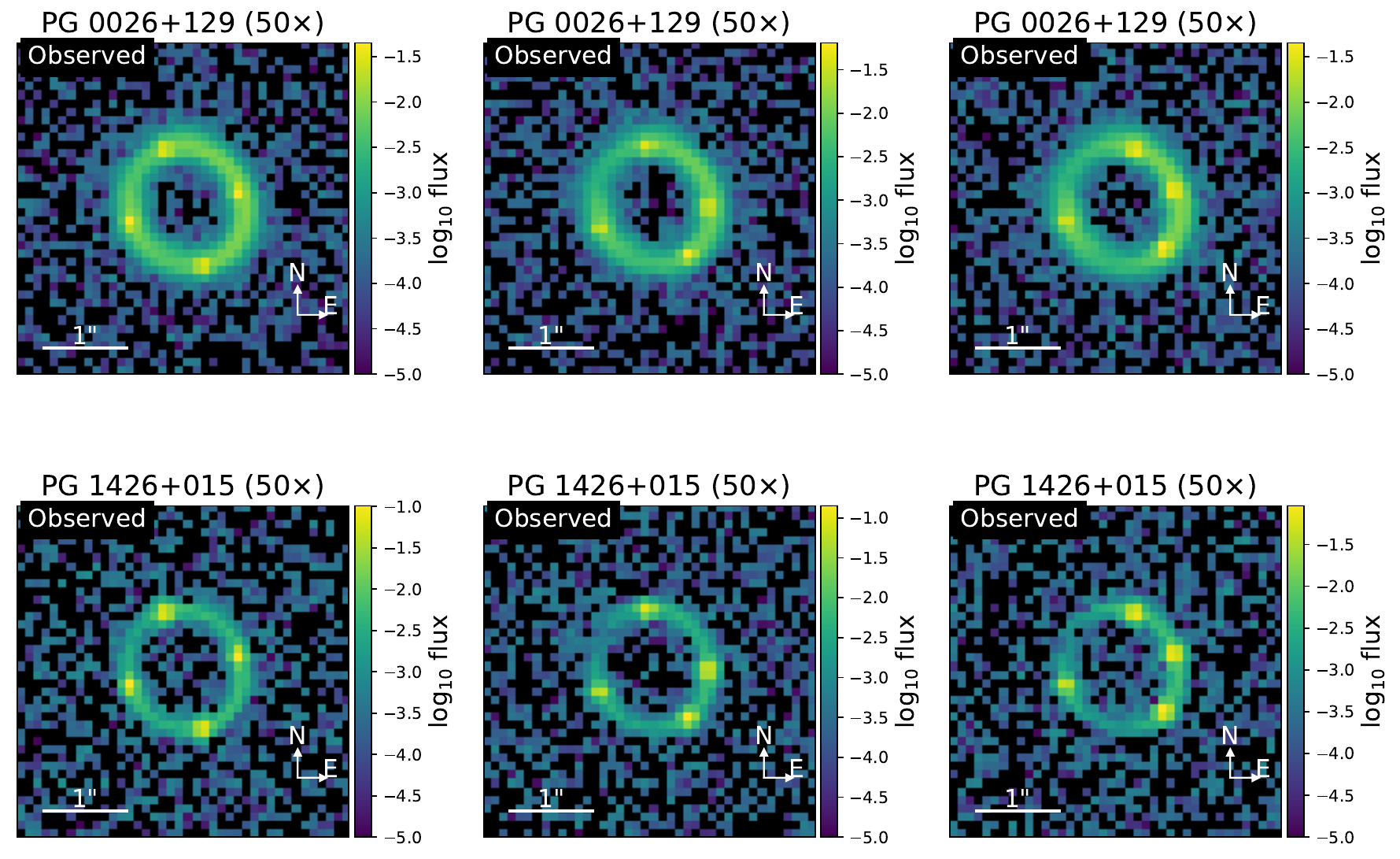}
    \caption{Mock lenses for PG~0026+129 and PG~1426+015 with extended [O\,III] emission boosted 50$\times$ in cross (left), fold (center), and cusp (right) configurations. Both show detectable extended emission by-eye inspection, requiring our PS + Extended Source model.}
    \label{fig:50XLenses}
\end{figure}

\begin{figure}
    \includegraphics[width=1\linewidth]{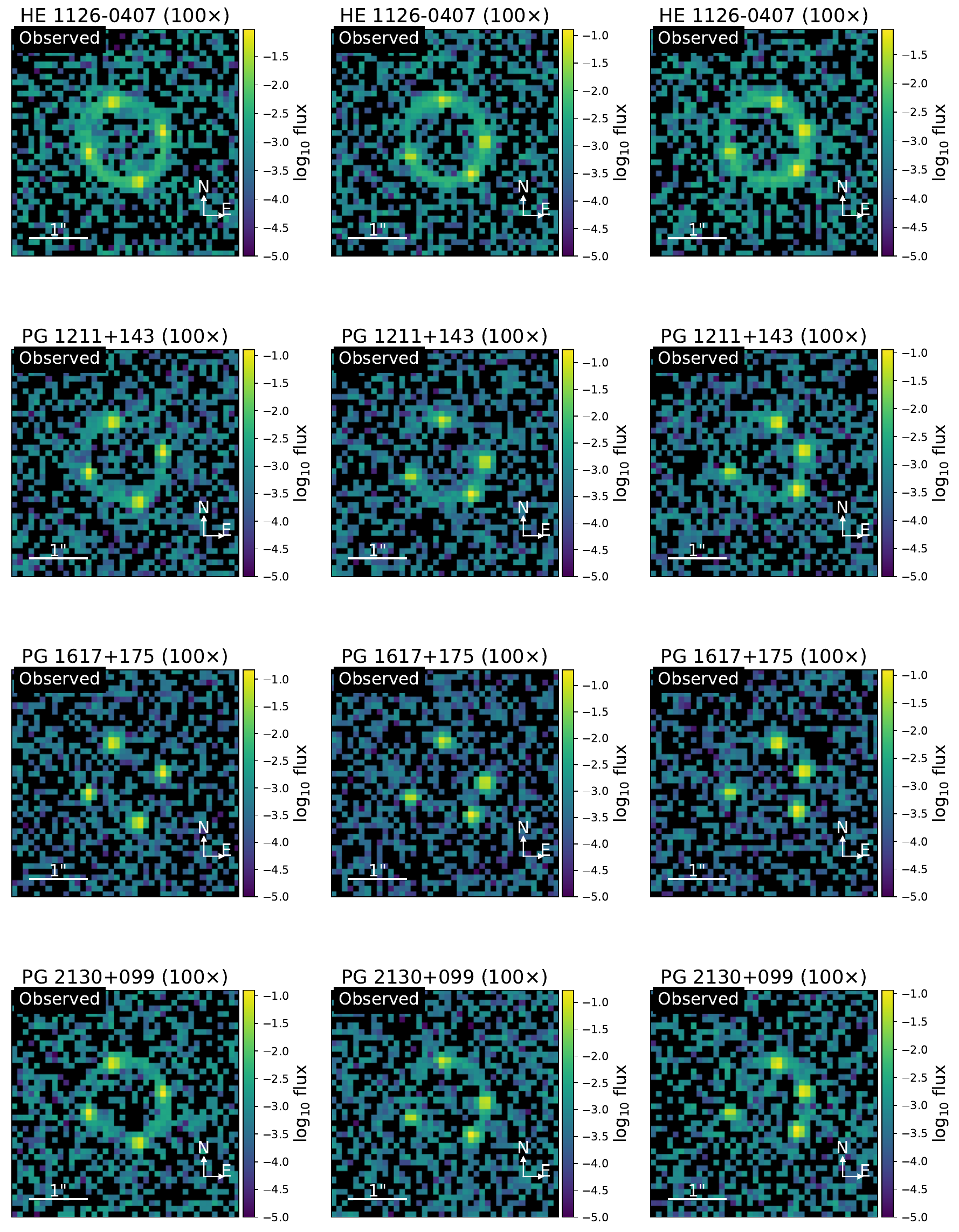}
    \caption{Mock lenses for HE~1126-0407, PG~1211+143, PG~1617+175, and PG~2130+099 with extended [O\,III] emission boosted 100$\times$ in cross (left), fold (center), and cusp (right) configurations. Extended emission detectable by-eye in three (PS + Extended Source model required); PG~1617+175 remains point-source only.}
    \label{fig:100XLenses}
\end{figure}

To probe extended [O\,III] detectability limits and analyze a wider variety of scenarios, we generate additional simulations scaling extended [O\,III] surface brightness by 50$\times$ or 100$\times$ relative to the original cases (Figs.~\ref{fig:50XLenses}--\ref{fig:100XLenses}). These 50$\times$/100$\times$ scalings span realistic parameter space extremes, testing robustness beyond the expected [O\,III] brightness ratios. The white-light images (narrow-band around [O\,III] $\lambda5007$) reveal by-eye detection of extended emission in all but PG~1617+175 (even at 100$\times$, due to compact/faint source morphology). We apply identical fitting and spectral extraction procedures to assess impact on point-source spectra and flux ratios.

\section{Results}\label{sec:results}

\begin{figure}
    \centering
    \includegraphics[width=1\linewidth]{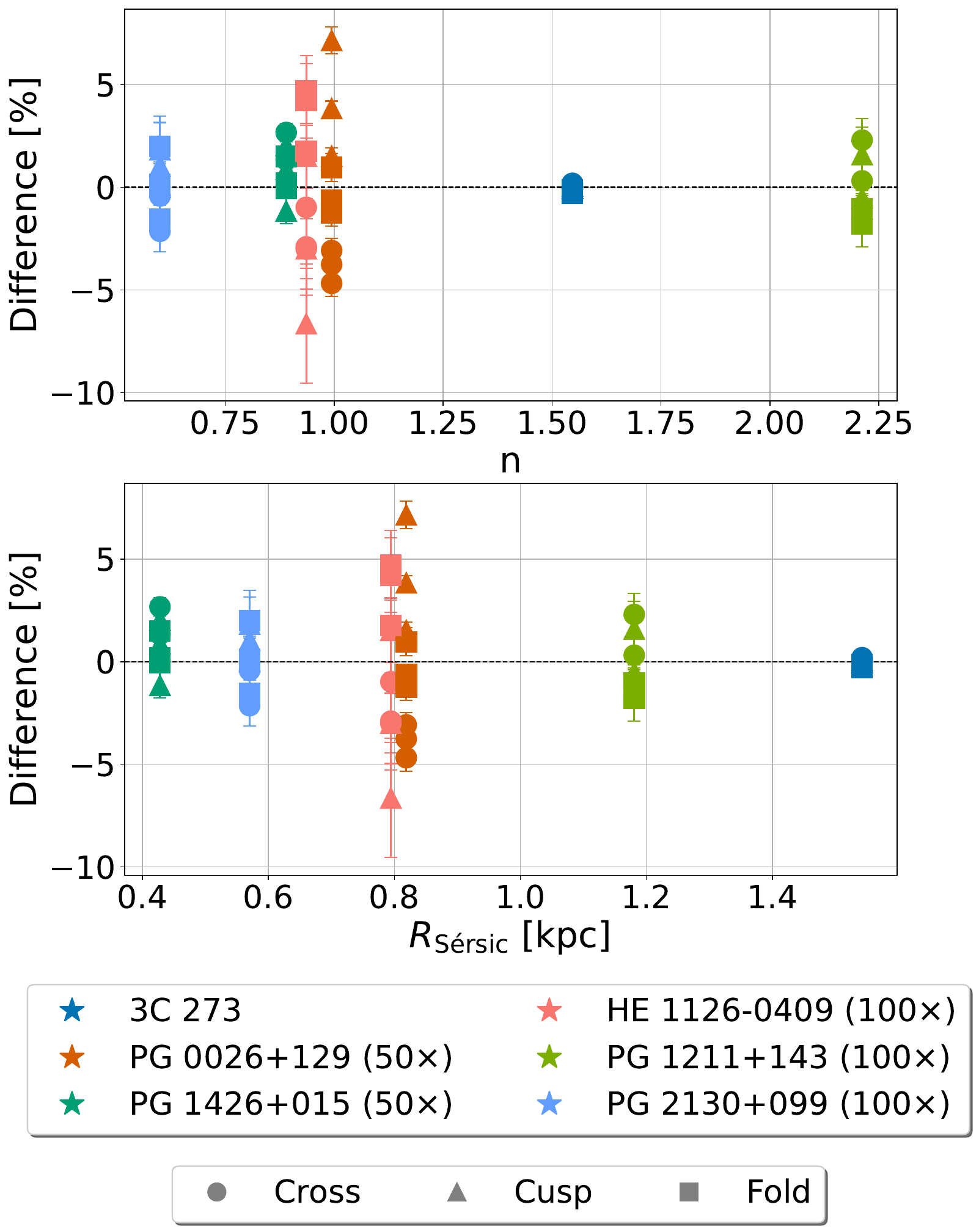}
    \caption{Percent differences between true/recovered flux ratios (B/A, C/A, D/A) vs.\ S\'ersic index $n$ (top) and S\'ersic radius $R_\mathrm{\text{S\'ersic}}$ (bottom). Symbols correspond to lensing configurations: cross ($\circ$), cusp ($\triangle$), fold ($\square$). Three points per x-value per configuration correspond to the three image flux ratios (B/A, C/A, D/A) for six sources using PS + Extended Source model. Flux ratio uncertainties $<5\%$, well below typical $\sim6\%$ spectral fitting baseline \cite{Nierenberg2019}.}
    \label{fig:fluxRatios}
\end{figure}

\begin{figure}
    \centering
    \includegraphics[width=1\linewidth]{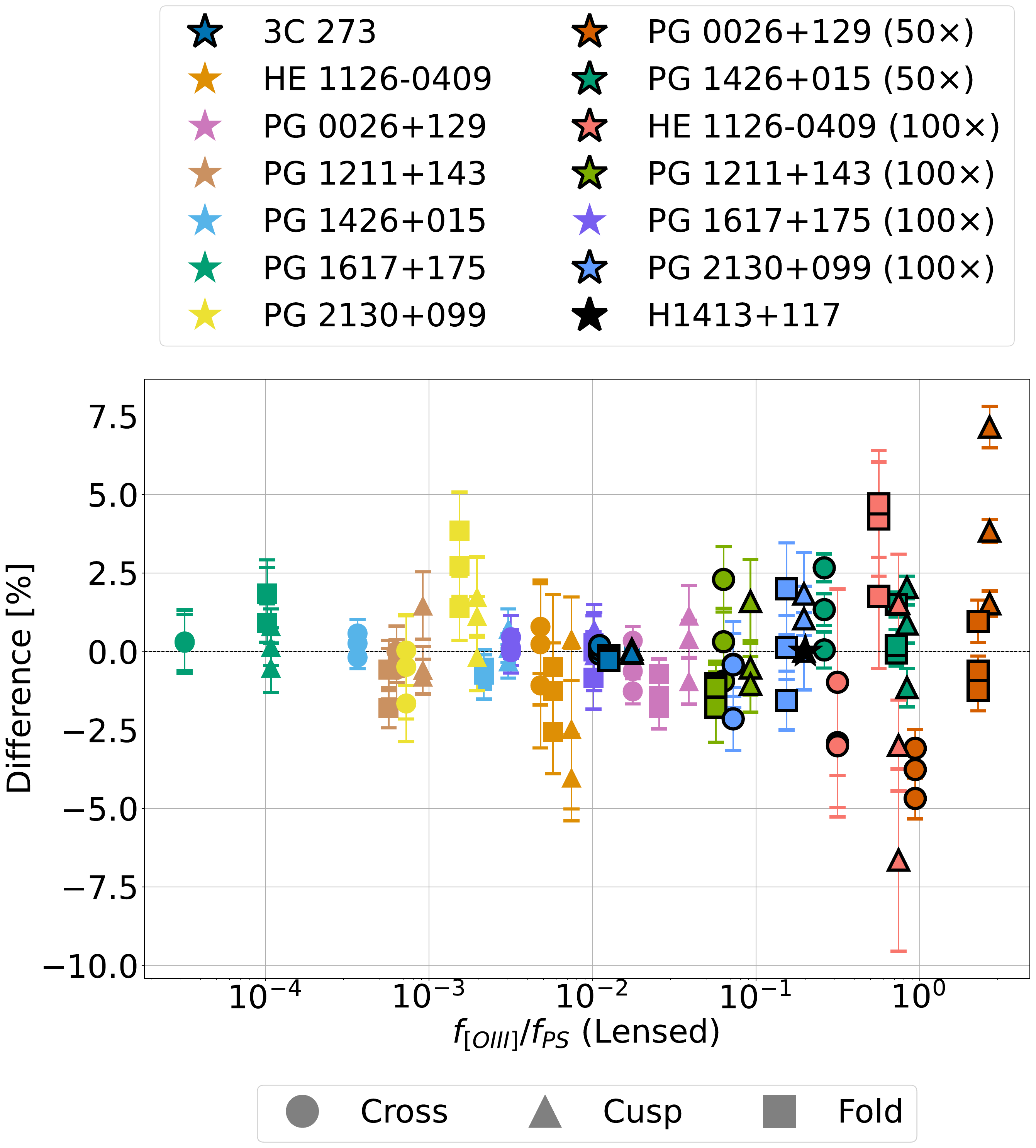}
    \caption{Percent differences between true/recovered {\it lensed flux ratios} (B/A, C/A, D/A for DM studies) vs.\ local lensed [O\,III]/point-source flux ratio ($f_{[O\,III]}/f_\mathrm{PS}$) within $0\farcs3\times0\farcs3$ apertures around model point-source positions. Symbols correspond to lensing configurations: cross ($\circ$), cusp ($\triangle$), fold ($\square$). Black-outlined points use PS + Extended Source model (extended [O\,III] detected by-eye); others use point-source-only. Black star: H~1413+117 ``Cloverleaf" quasar lens with real extended [O\,III] (Full analysis in Nierenberg et al.\ in prep.). Detection threshold for PS + Extended Source modeling: $f_{[O\,III]}/f_\mathrm{PS} \sim 0.01$. Flux ratio uncertainties $<5\%$ for $f_{[O\,III]}/f_\mathrm{PS} < 1$, below typical $\sim6\%$ spectral fitting baseline \cite{Nierenberg2019}.}
    \label{fig:moneyPlot}
\end{figure}

In this section, we present our key findings on flux-ratio recovery accuracy. We compute recovered flux ratios from extracted spectra where extended emission is present. Flux ratios---key observables for substructure and dark-matter studies---are obtained by integrating input and recovered spectra over the redshifted [O\,III] $\lambda5007$ range, then taking ratios (B/A, C/A, D/A) relative to the brightest image A (``true'' for input, ``recovered'' from extraction). Typical flux-ratio measurement uncertainty is $\sim6\%$ \cite{Nierenberg2019}; we use this as a benchmark to assess whether our PS + Extended Source modeling introduces significant additional uncertainty. The percent differences between the ``true" and ``recovered" ratios therefore directly measure the impact of extended [O\,III] emission relative to this baseline.

We investigate source morphology and lensing configuration effects in all simulations by plotting percent differences in recovered flux ratios versus S\'ersic index $n$ and S\'ersic radius $R_\mathrm{\text{S\'ersic}}$, (Fig.~\ref{fig:fluxRatios}). No systematic biases appear across varying configurations. Neither $n$ or $R_\mathrm{\text{S\'ersic}}$  correlates with percent differences , indicating the secondary role of morphology. Critically, simulations where extended emission is visually prominent show most flux ratios falling below 5\% difference---well within spectral fitting uncertainties.

To quantify possible extended [O\,III] contamination in flux-ratio recovery, we used lensed [O\,III]-only and point-source-only cubes to measure integrated fluxes within $0\farcs3\times0\farcs3$ apertures at model point-source positions (all simulations + boosted cases). The resulting local flux ratio $f_{[O\,III]}/f_\mathrm{PS}$ (x-axis, Fig.~\ref{fig:moneyPlot})---distinct from lensed flux ratios B/A, C/A, D/A (y-axis) used for dark matter inference---characterizes the relative extended emission near the images. The quadruply imaged quasar H~1413+117 (the “Cloverleaf” lens) provides an example of a real system in the regime where PS + Extended Source modeling is required, for which a full analysis is ongoing (Nierenberg et al.\ in prep.).
 The plot shows percent differences versus $f_{[O\,III]}/f_\mathrm{PS}$: PS + Extended Source fits are required for black-outlined points (extended [O\,III] detected by-eye in the lensed images); detection requires [O\,III] contributing $\gtrsim 1\%$ of the local point-source flux. PG~0026+129 falls below this threshold despite the intrinsically brighter [O\,III] emission due to low S/N.

Extended emission becomes detectable by-eye for $L_\mathrm{bol}\sim10^{46}$\,erg\,s$^{-1}$ quasars when $f_{[O\,III]}/f_\mathrm{PS}\sim0.01$--$0.1$. Critically, both point-source-only and two-component modeling maintain flux-ratio accuracy below $\sim6\%$ spectral uncertainties across this range -- well below uncertainties from [O\,III] disentanglement amid overlapping broad Fe\,{\sc ii}, continuum, and H$\beta$ emission during spectral fitting.

\section{Summary and Conclusions}\label{sec:sum&Conc}

In this paper, we present a flexible PS + Extended Source modeling framework to robustly handle potential unresolved extended [O\,III] emission in gravitationally lensed quasars observed with adaptive optics IFUs. Using realistic mock Keck OSIRIS AO observations, we created a pipeline that simultaneously fits nuclear point sources simultaneously with extended emission modeled as S\'ersic ellipses, demonstrating sufficient accuracy for robust flux-ratio recovery even when extended emission is present or absent.

We span representative cross, cusp, and fold lensing configurations using high-resolution [O\,III] emission maps from the CARS survey \cite{Husemann2022} and Winkel25, redshifted into OSIRIS H-band ($1.47$--$1.80$\,$\mu$m). Lensed images simulated with \texttt{lenstronomy} SIE+external shear macromodels incorporate realistic OSIRIS AO sky noise, telluric features, and seeing-limited PSF convolution across varied source morphologies ($n=0.5$--$4$, $R_\mathrm{eff}=0.5$--$2$\,kpc).

Our modeling pipeline employs hybrid particle swarm optimization with multi-start initialization to decompose lensed images into PS + Extended Source components, optimizing Sérsic parameters ($n$, $R_\mathrm{eff}$, ellipticity, position angle) plus point-source fluxes constrained by macromodel positions.

\textbf{Key findings}:
\begin{itemize}
\item Flux-ratio recovery does not show dependence on extended source morphology ($n$, $R_\mathrm{eff}$) or lensing configuration; most B/A, C/A, D/A ratios recover to $<5\%$ across parameter space (Fig.~\ref{fig:fluxRatios})
\item PS + Extended Source modeling maintains $<5\%$ accuracy even when $f_{[O\,III]}/f_\mathrm{PS} \sim 0.01$--$0.1$ (visually prominent emission); the PS-only model performs equivalently below the detection threshold $\sim0.01$ (Fig.~\ref{fig:moneyPlot})
\item Visual inspection of white-light images reliably determines the appropriate modeling choice
\item All uncertainties remain below the typical $\sim6\%$ spectral fitting baseline from broad-line decomposition \cite{Nierenberg2019}
\item S\'ersic elliptical profiles sufficiently capture the extended [O\,III] morphology for accurate point-source flux isolation
\end{itemize}

These results have direct implications for dark matter substructure searches via flux-ratio anomalies in quadruply imaged quasars---one of the cleanest astrophysical probes of milli-lensing by low-mass halos. With thousands of new strong lenses expected from LSST, Euclid, and Rubin Observatory, our validated pipeline ensures systematic, robust flux-ratio measurements regardless of NLR properties, applicable to existing OSIRIS AO/NIFS datasets and upcoming JWST/NIRSpec integral-field-unit observations.

\section*{Acknowledgments}

We thank Nico Winkel for invaluable assistance with data management, processing, and providing access to the datasets essential for this analysis. VNB acknowledges support from the National Science Foundation under grants AST-1909297. AMN and MFPM acknowledge support from National Science Foundation through the grant “Collaborative Research: Measuring the physical properties of dark
matter with strong gravitational lensing"
NSF-AST-2206315, and from the
the National Science Foundation through the grant “CAREER:
An order of magnitude improvement in measurements of the
physical properties of dark matter" NSF-AST-2442975.

Note that findings and conclusions do not necessarily represent views of the NSF. 
\\



\appendix*
\section{Additional two-component lens modeling results}\label{app:twoCompFits}

In this appendix, we present the full set of two-component (PS + extended source) lens modeling results for the simulated systems HE~1126$-$0407, PG~1211+143, PG~2130+099, PG~0026+129, and PG~1426+015 at enhanced [O\,III] surface brightness levels (50$\times$ or 100$\times$). 
For each source, we show the cross, cusp, and fold image configurations in separate rows, and organize the panels such that the simulated lensed data, best-fit model reconstruction, and normalized residuals occupy the left, middle, and right columns, respectively.

Visual inspection of the residual maps demonstrates that the two-component source model provides an excellent description of the simulated data across all systems and configurations. 
The normalized residuals are generally low-amplitude and lack coherent spatial structure, indicating that both the compact AGN point source and the extended NLR emission are well captured by the model. 
In the case of PG~0026+129, the cross configuration exhibits a slight apparent tilt or asymmetry in the residuals. Importantly, even in this case, the deviations remain small and do not significantly impact the recovered flux ratios. Overall, these results confirm that the two-component modeling approach remains robust even in regimes with relatively bright extended [O\,III] emission. 
The fits shown here complement the quantitative flux-ratio analysis in the main text and support the conclusion that accurate point-source fluxes can be recovered in the presence of detectable NLR structure, for both moderate (50$\times$) and more extreme (100$\times$) [O\,III] enhancements.

\begin{figure}
    \centering
    \includegraphics[width=1\linewidth]{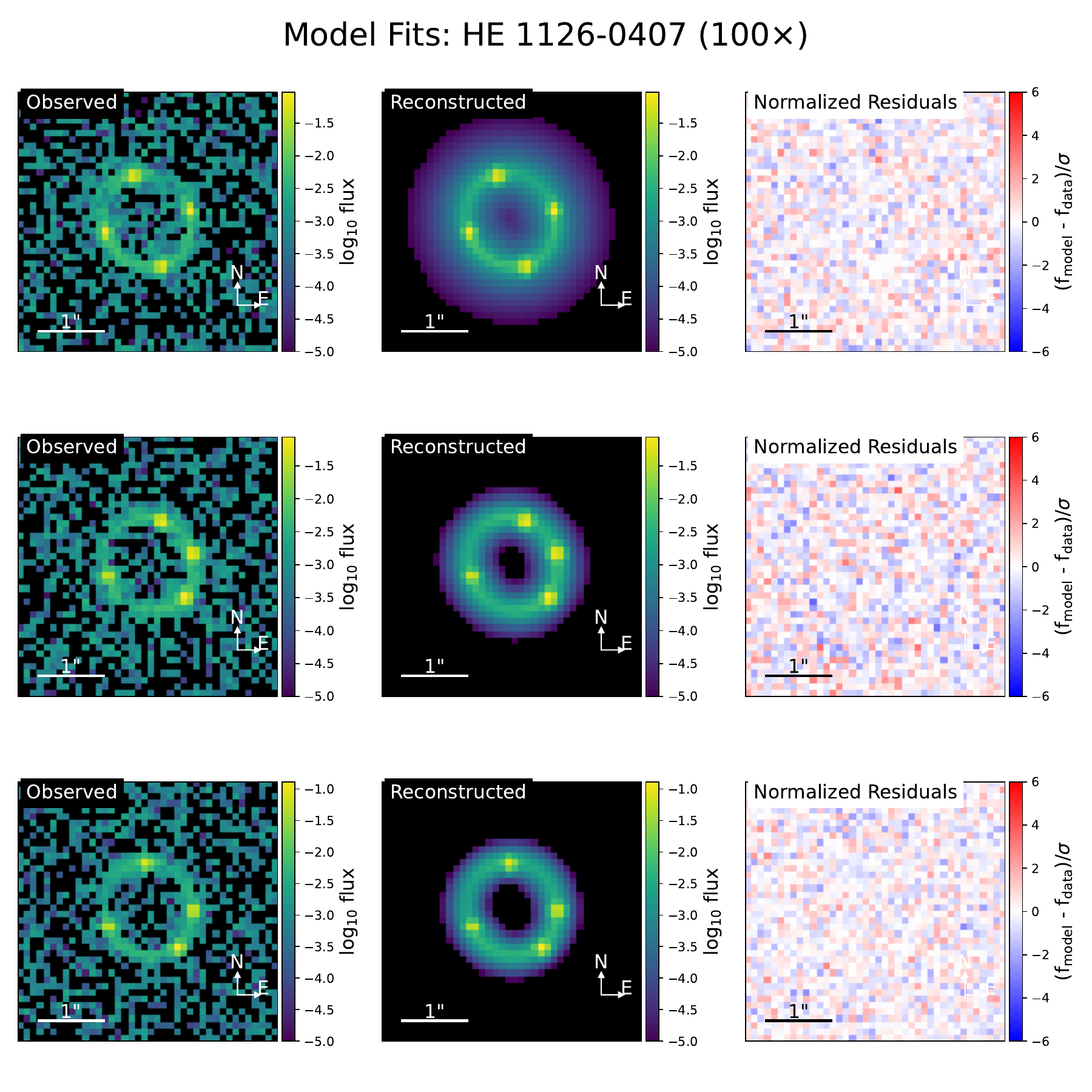}
    \caption{Two-component (PS + extended source) lens modeling results for HE~1126-0407 (100$\times$). Each row corresponds to a different image configuration: cross (top), cusp (middle), and fold (bottom). The left column shows the simulated lensed [O\,III] emission, the middle column shows the best-fit model reconstruction, and the right column shows the normalized residuals. In all configurations, the residuals are consistent with noise, indicating that the two-component model provides an excellent description of both the compact and extended emission.}
    \label{fig:he1126_100xFit}
\end{figure}

\begin{figure}
    \centering
    \includegraphics[width=1\linewidth]{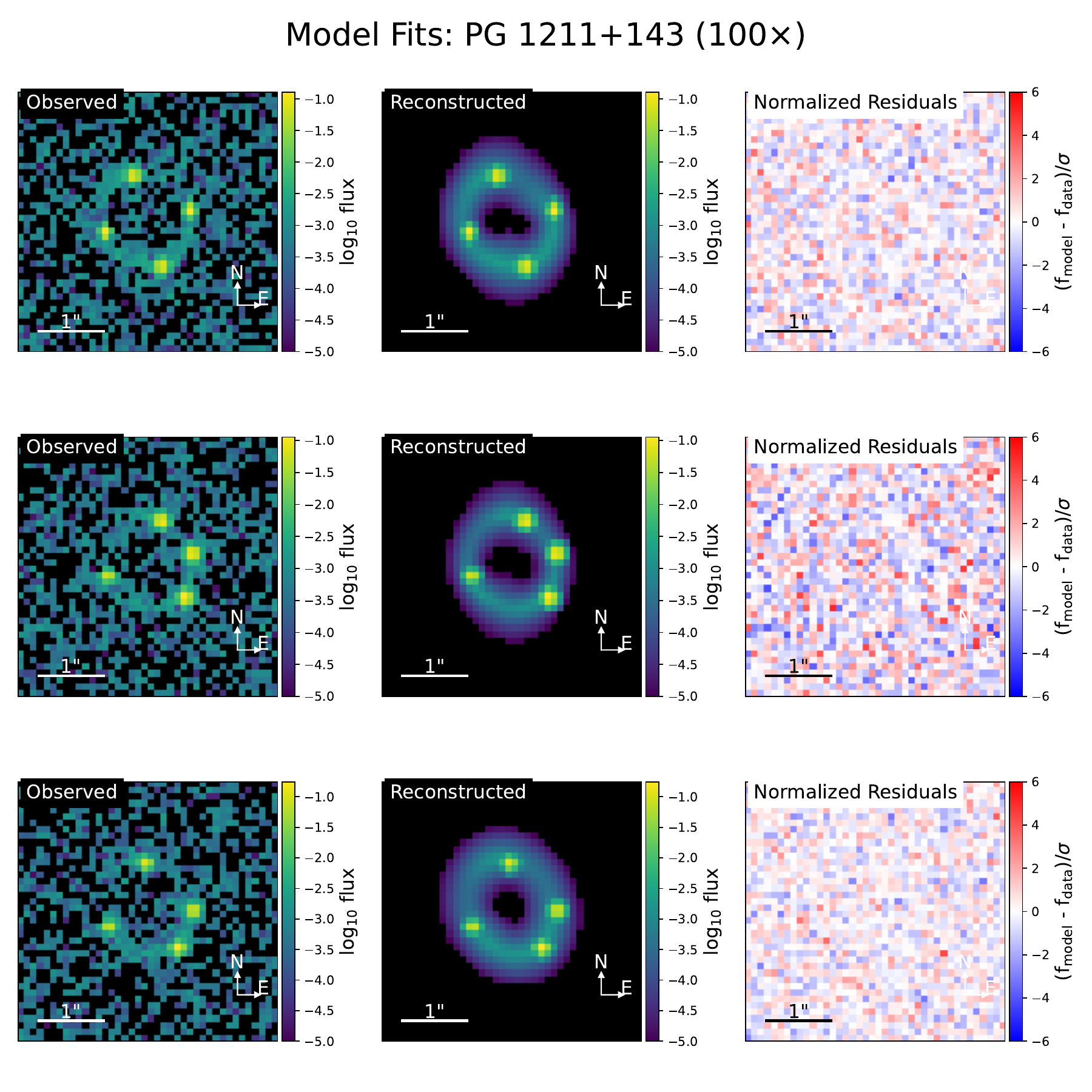}
    \caption{Two-component (PS + extended source) lens modeling results for PG~1211+143 (100$\times$). As in Fig.~\ref{fig:he1126_100xFit}, rows indicate the cross, cusp, and fold configurations, while columns show the simulated lensed data (left), model reconstruction (middle), and normalized residuals (right). The low-amplitude, structureless residuals across all configurations demonstrate that the extended NLR component and the point-source emission are both well captured by the model.}
    \label{fig:pg1211_100xFit}
\end{figure}
\begin{figure}
    \centering
    \includegraphics[width=1\linewidth]{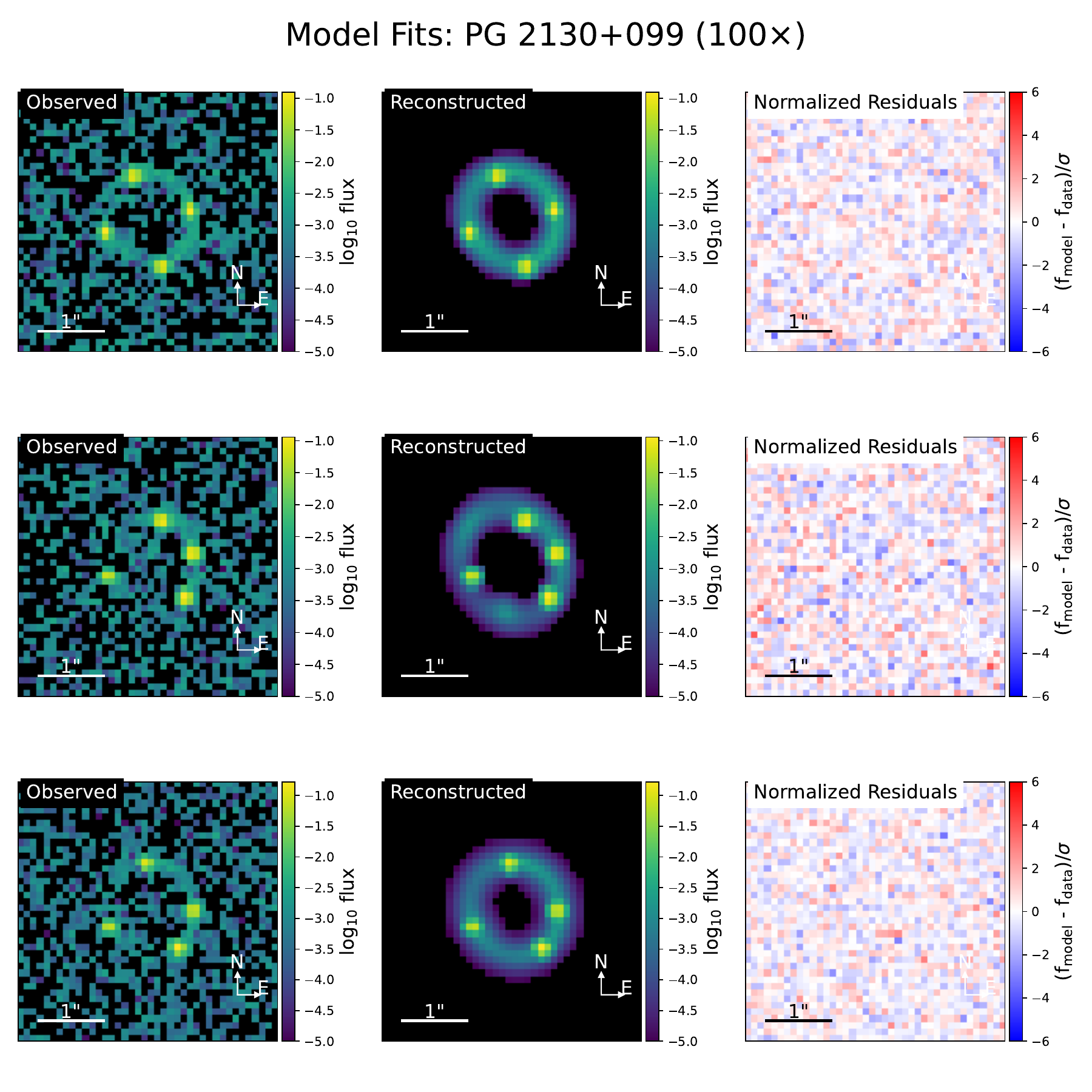}
    \caption{Two-component (PS + extended source) lens modeling results for PG~2130+099 (100$\times$). 
    The cross, cusp, and fold configurations are shown from top to bottom, with the simulated lensed images, model reconstructions, and normalized residuals displayed from left to right. 
    The absence of coherent residual structures indicates that the inferred source model provides a statistically acceptable fit to the simulated data in all image configurations.}
    \label{fig:pg2130_100xFit}
\end{figure}
\begin{figure}
    \centering
    \includegraphics[width=1\linewidth]{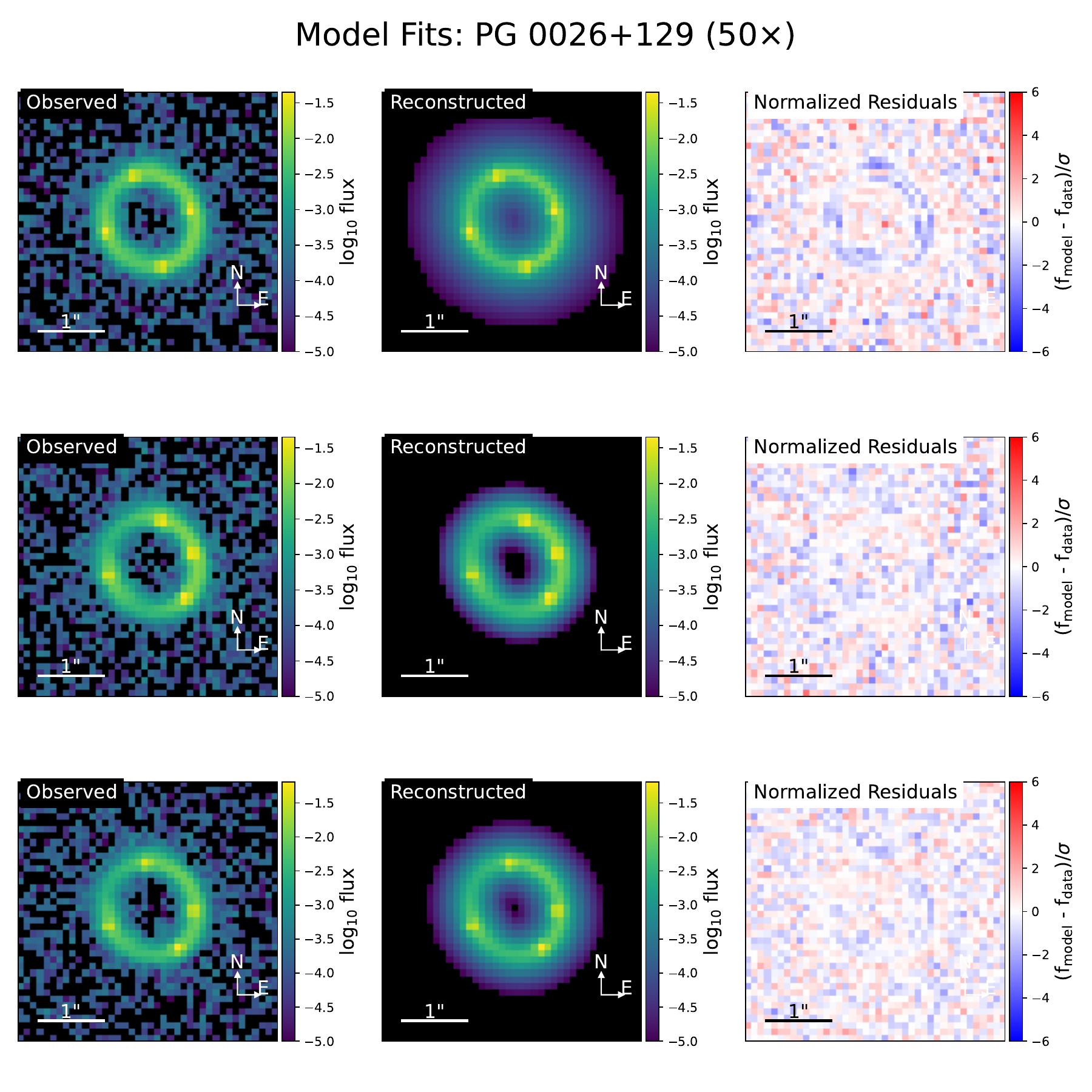}
    \caption{Two-component (PS + extended source) lens modeling results for PG~0026+129 (50 $\times $). As in the previous figures, the cross, cusp, and fold configurations are shown from top to bottom, with the simulated lensed images, model reconstructions, and normalized residuals in the left, middle, and right columns, respectively. The cross configuration exhibits a slight asymmetry in the residuals, which arises from the interplay between the extended emission morphology and the lensing caustic geometry, but the overall residuals remain small and the fits are consistent with a good reconstruction of both the point-like and extended components.}
    \label{fig:pg0026_50xFit}
\end{figure}
\begin{figure}
    \centering
    \includegraphics[width=1\linewidth]{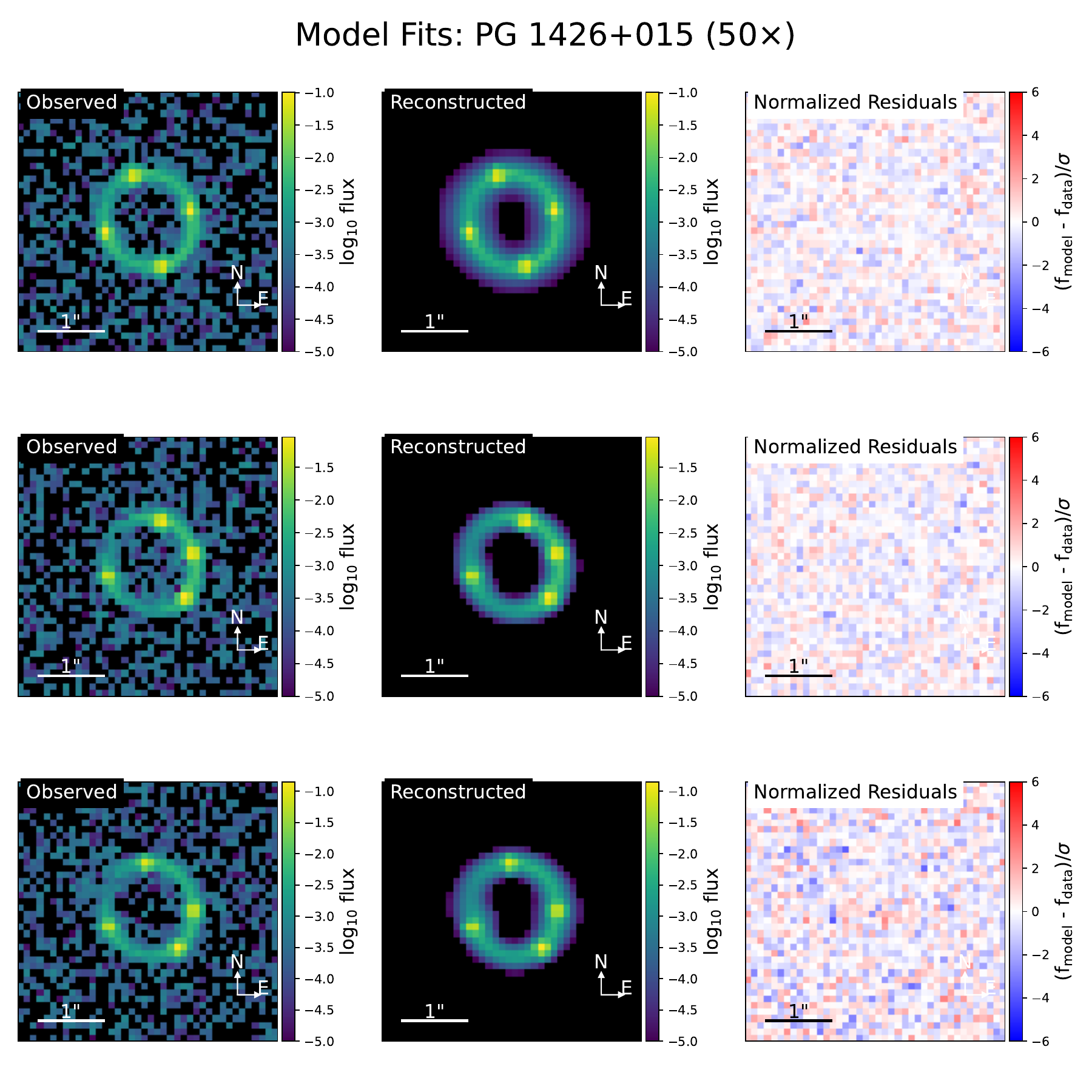}
    \caption{Two-component (PS + extended source) lens modeling results for PG~1426+015 (50$\times$). From top to bottom, the rows correspond to the cross, cusp, and fold configurations, while the columns show the simulated lensed images, model reconstructions, and normalized residuals. The residual maps do not show significant structured deviations, confirming that the adopted two-component source model provides a robust description of the simulated lensed [O\,III] emission for this system.}
    \label{fig:pg1426_50xFit}
\end{figure}

\nocite{*}

\bibliography{references}

@article{review_of_part_22,
	title = {Review of particle physics},
	volume = {2022},
	issn = {2050-3911},
	url = {https://doi.org/10.1093/ptep/ptac097},
	doi = {10.1093/ptep/ptac097},
	number = {8},
	journal = {Progress of Theoretical and Experimental Physics},
	author = {Group, Particle Data and Workman, R L and Burkert, V D and Crede, V and Klempt, E and Thoma, U and Tiator, L and Agashe},
	month = aug,
	year = {2022},
	note = {tex.eprint: https://academic.oup.com/ptep/article-pdf/2022/8/083C01/49175539/ptac097.pdf},
	pages = {083C01},
}

@article{gilman_constraining_2023,
  title = {Constraining Resonant Dark Matter Self-Interactions with Strong Gravitational Lenses},
  author = {Gilman, Daniel and Zhong, Yi-Ming and Bovy, Jo},
  year = 2023,
  journal = {Phys. Rev. D},
  volume = {107},
  number = {10},
  pages = {103008},
  doi = {10.1103/PhysRevD.107.103008}
}

@article{gilman_constraints_2020,
  title = {Constraints on the Mass-Concentration Relation of Cold Dark Matter Halos with 11 Strong Gravitational Lenses},
  author = {Gilman, Daniel and Du, Xiaolong and Benson, Andrew and Birrer, Simon and Nierenberg, Anna and Treu, Tommaso},
  year = 2020,
  journal = {MNRAS},
  volume = {492},
  pages = {L12--L16},
  issn = {0035-8711},
  doi = {10.1093/mnrasl/slz173}
}

@article{laroche_quantum_2022,
  title = {Quantum Fluctuations Masquerade as Haloes: Bounds on Ultra-Light Dark Matter from Quadruply Imaged Quasars},
  shorttitle = {Quantum Fluctuations Masquerade as Haloes},
  author = {Laroche, Alexander and Gilman, Daniel and Li, Xinyu and Bovy, Jo and Du, Xiaolong},
  year = 2022,
  journal = {MNRAS},
  volume = {517},
  pages = {1867--1883},
  issn = {0035-8711},
  doi = {10.1093/mnras/stac2677}
}

@article{gilman_primordial_2022,
  title = {The Primordial Matter Power Spectrum on Sub-Galactic Scales},
  author = {Gilman, Daniel and Benson, Andrew and Bovy, Jo and Birrer, Simon and Treu, Tommaso and Nierenberg, Anna},
  year = 2022,
  journal = {MNRAS},
  volume = {512},
  number = {3},
  pages = {3163--3188},
  doi = {10.1093/mnras/stac670}
}

@Article{Granato1996,
  author     = {Granato, Gian Luigi and Danese, Luigi and Franceschini, Alberto},
  journal    = {The Astrophysical Journal},
  title      = {Dust-Enshrouded AGN Models for Hyperluminous, High-Redshift Infrared Galaxies},
  year       = {1996},
  issn       = {0004-637X},
  month      = mar,
  number     = {1},
  volume     = {460},
  doi        = {10.1086/309977},
  publisher  = {American Astronomical Society},
  readstatus = {skimmed},
}

@Article{Nierenberg2019,
  author    = {Nierenberg, A M and Gilman, D and Treu, T and Brammer, G and Birrer, S and Moustakas, L and Agnello, A and Anguita, T and Fassnacht, C D and Motta, V and Peter, A H G and Sluse, D},
  journal   = {Monthly Notices of the Royal Astronomical Society},
  title     = {Double dark matter vision: twice the number of compact-source lenses with narrow-line lensing and the WFC3 grism},
  year      = {2019},
  issn      = {1365-2966},
  month     = dec,
  number    = {4},
  pages     = {5314--5335},
  volume    = {492},
  doi       = {10.1093/mnras/stz3588},
  publisher = {Oxford University Press (OUP)},
}

@Article{Moustakas2003,
  author    = {Moustakas, L. A. and Metcalf, R. B.},
  journal   = {Monthly Notices of the Royal Astronomical Society},
  title     = {Detecting dark matter substructure spectroscopically in strong gravitational lenses},
  year      = {2003},
  issn      = {1365-2966},
  month     = mar,
  number    = {3},
  pages     = {607--615},
  volume    = {339},
  doi       = {10.1046/j.1365-8711.2003.06055.x},
  publisher = {Oxford University Press (OUP)},
}

@Article{Nierenberg2017,
  author    = {Nierenberg, A. M. and Treu, T. and Brammer, G. and Peter, A. H. G. and Fassnacht, C. D. and Keeton, C. R. and Kochanek, C. S. and Schmidt, K. B. and Sluse, D. and Wright, S. A.},
  journal   = {Monthly Notices of the Royal Astronomical Society},
  title     = {Probing dark matter substructure in the gravitational lens HE 0435-1223 with the WFC3 grism},
  year      = {2017},
  issn      = {1365-2966},
  month     = jun,
  number    = {2},
  pages     = {2224--2236},
  volume    = {471},
  doi       = {10.1093/mnras/stx1400},
  publisher = {Oxford University Press (OUP)},
}

@Article{Birrer2018,
  author    = {Birrer, Simon and Amara, Adam},
  journal   = {Physics of the Dark Universe},
  title     = {lenstronomy: Multi-purpose gravitational lens modelling software package},
  year      = {2018},
  issn      = {2212-6864},
  month     = dec,
  pages     = {189--201},
  volume    = {22},
  doi       = {10.1016/j.dark.2018.11.002},
  publisher = {Elsevier BV},
}

@Article{Bennert2002,
  author    = {Bennert, Nicola and Falcke, Heino and Schulz, Hartmut and Wilson, Andrew S. and Wills, Beverley J.},
  journal   = {The Astrophysical Journal},
  title     = {Size and Structure of the Narrow-Line Region of Quasars},
  year      = {2002},
  issn      = {1538-4357},
  month     = aug,
  number    = {2},
  pages     = {L105--L109},
  volume    = {574},
  doi       = {10.1086/342420},
  publisher = {American Astronomical Society},
}

@Article{Husemann2022,
  author    = {Husemann, B. and Singha, M. and Scharwächter, J. and McElroy, R. and Neumann, J. and Smirnova-Pinchukova, I. and Urrutia, T. and Baum, S. A. and Bennert, V. N. and Combes, F. and Croom, S. M. and Davis, T. A. and Fournier, Y. and Galkin, A. and Gaspari, M. and Enke, H. and Krumpe, M. and O’Dea, C. P. and Pérez-Torres, M. and Rose, T. and Tremblay, G. R. and Walcher, C. J.},
  journal   = {Astronomy \& Astrophysics},
  title     = {The Close AGN Reference Survey (CARS): IFU survey data and the BH mass dependence of long-term AGN variability},
  year      = {2022},
  issn      = {1432-0746},
  month     = mar,
  pages     = {A124},
  volume    = {659},
  doi       = {10.1051/0004-6361/202141312},
  publisher = {EDP Sciences},
}

@Article{Stockton1987,
  author     = {Stockton, Alan and MacKenty, John W.},
  journal    = {The Astrophysical Journal},
  title      = {Extended emission-line regions around QSOs},
  year       = {1987},
  issn       = {1538-4357},
  month      = may,
  pages      = {584},
  volume     = {316},
  doi        = {10.1086/165227},
  publisher  = {American Astronomical Society},
  readstatus = {skimmed},
}

@Article{Husemann2012,
  author     = {Husemann, B. and Wisotzki, L. and Sánchez, S. F. and Jahnke, K.},
  journal    = {Astronomy \& Astrophysics},
  title      = {The properties of the extended warm ionised gas around low-redshift QSOs and the lack of extended high-velocity outflows},
  year       = {2012},
  issn       = {1432-0746},
  month      = dec,
  pages      = {A43},
  volume     = {549},
  doi        = {10.1051/0004-6361/201220076},
  publisher  = {EDP Sciences},
  readstatus = {skimmed},
}

@Article{IbarraMedel2024,
  author     = {Ibarra-Medel, H and Negrete, C A and Lacerna, I and Hernández-Toledo, H M and Cortes-Suárez, E and Sánchez, S F},
  journal    = {Monthly Notices of the Royal Astronomical Society},
  title      = {An iterative method to deblend AGN-Host contributions for Integral Field spectroscopic observations},
  year       = {2024},
  issn       = {1365-2966},
  month      = nov,
  number     = {1},
  pages      = {752--776},
  volume     = {536},
  doi        = {10.1093/mnras/stae2623},
  publisher  = {Oxford University Press (OUP)},
  readstatus = {skimmed},
}

@Article{Chen2019,
  author     = {Chen, Jianhang and Shi, Yong and Dempsey, Ross and Law, David R and Chen, Yanmei and Yan, Renbin and Bing, Longji and Rembold, Sandro B and Li, Songlin and Yu, Xiaoling and Riffel, Rogemar A and Brownstein, Joe R and Riffel, Rogério},
  journal    = {Monthly Notices of the Royal Astronomical Society},
  title      = {The spatial extension of extended narrow line regions in MaNGA AGN},
  year       = {2019},
  issn       = {1365-2966},
  month      = aug,
  number     = {1},
  pages      = {855--867},
  volume     = {489},
  doi        = {10.1093/mnras/stz2183},
  publisher  = {Oxford University Press (OUP)},
  readstatus = {skimmed},
}

@Misc{Hogg1999,
  author    = {Hogg, David W.},
  title     = {Distance measures in cosmology},
  year      = {1999},
  copyright = {Assumed arXiv.org perpetual, non-exclusive license to distribute this article for submissions made before January 2004},
  doi       = {10.48550/ARXIV.ASTRO-PH/9905116},
  publisher = {arXiv},
}

@Article{Bennert2006,
  author    = {Bennert, N. and Jungwiert, B. and Komossa, S. and Haas, M. and Chini, R.},
  journal   = {Astronomy \& Astrophysics},
  title     = {Size and properties of the narrow-line region in Seyfert-2 galaxies from spatially-resolved optical spectroscopy},
  year      = {2006},
  issn      = {1432-0746},
  month     = sep,
  number    = {3},
  pages     = {953--966},
  volume    = {456},
  doi       = {10.1051/0004-6361:20065319},
  publisher = {EDP Sciences},
}

@Misc{Vegetti2023,
  author    = {Vegetti, S. and Birrer, S. and Despali, G. and Fassnacht, C. D. and Gilman, D. and Hezaveh, Y. and Levasseur, L. Perreault and McKean, J. P. and Powell, D. M. and O'Riordan, C. M. and Vernardos, G.},
  title     = {Strong gravitational lensing as a probe of dark matter},
  year      = {2023},
  copyright = {arXiv.org perpetual, non-exclusive license},
  doi       = {10.48550/ARXIV.2306.11781},
  publisher = {arXiv},
}

@Article{Winkel2025,
  author    = {Winkel, Nico and Bennert, Vardha N. and Remigio, Raymond P. and Treu, Tommaso and Jahnke, Knud and U, Vivian and Barth, Aaron J. and Malkan, Matthew and Husemann, Bernd and Ding, Xuheng and Birrer, Simon},
  journal   = {The Astrophysical Journal},
  title     = {Combining Direct Black Hole Mass Measurements and Spatially Resolved Stellar Kinematics to Calibrate the $M_\mathrm{BH}-\sigma_\star$ Relation of Active Galaxies},
  year      = {2025},
  issn      = {1538-4357},
  month     = dec,
  number    = {1},
  pages     = {115},
  volume    = {978},
  doi       = {10.3847/1538-4357/ad9272},
  publisher = {American Astronomical Society},
}

@InProceedings{Bacon2010,
  author    = {Bacon, R. and Accardo, M. and Adjali, L. and Anwand, H. and Bauer, S. and Biswas, I. and Blaizot, J. and Boudon, D. and Brau-Nogue, S. and Brinchmann, J. and Caillier, P. and Capoani, L. and Carollo, C. M. and Contini, T. and Couderc, P. and Daguisé, E. and Deiries, S. and Delabre, B. and Dreizler, S. and Dubois, J. and Dupieux, M. and Dupuy, C. and Emsellem, E. and Fechner, T. and Fleischmann, A. and François, M. and Gallou, G. and Gharsa, T. and Glindemann, A. and Gojak, D. and Guiderdoni, B. and Hansali, G. and Hahn, T. and Jarno, A. and Kelz, A. and Koehler, C. and Kosmalski, J. and Laurent, F. and Le Floch, M. and Lilly, S. J. and Lizon, J.-L. and Loupias, M. and Manescau, A. and Monstein, C. and Nicklas, H. and Olaya, J.-C. and Pares, L. and Pasquini, L. and Pécontal-Rousset, A. and Pelló, R. and Petit, C. and Popow, E. and Reiss, R. and Remillieux, A. and Renault, E. and Roth, M. and Rupprecht, G. and Serre, D. and Schaye, J. and Soucail, G. and Steinmetz, M. and Streicher, O. and Stuik, R. and Valentin, H and Vernet, J. and Weilbacher, P. and Wisotzki, L. and Yerle, N.},
  booktitle = {Ground-based and Airborne Instrumentation for Astronomy III},
  title     = {The MUSE second-generation VLT instrument},
  year      = {2010},
  editor    = {McLean, Ian S. and Ramsay, Suzanne K. and Takami, Hideki},
  month     = jul,
  publisher = {SPIE},
  doi       = {10.1117/12.856027},
  issn      = {0277-786X},
}

@manual{osiris,
  title        = {OSIRIS User Manual},
  author       = {Larkin, James and Barczys, Matthew and McElwain, Mike and Perrin, Marshall and Weiss, Jason and Wright, Shelley and Lyke, Jim and Do, Tuan and Boehle, Anna and Chappell, Sam and Chu, Devin and Ciurlo, Anna and Fitzgerald, Michael and Lu, Jessica and Vayner, Andrey and Walth, Greg and Yeh, Sherry and Rundquist, Nils-Erik and Freeman, Matthew},
  year         = {2022},
  organization = {California Association for Research in Astronomy},
  edition      = {Version 6.0},
  note         = {UCLA Infrared Laboratory, Keck Observatory},
  url          = {https://www2.keck.hawaii.edu/inst/osiris/OSIRIS_Manual.pdf}
}

@Article{Nierenberg2014,
  author    = {Nierenberg, A. M. and Treu, T. and Wright, S. A. and Fassnacht, C. D. and Auger, M. W.},
  journal   = {Monthly Notices of the Royal Astronomical Society},
  title     = {Detection of substructure with adaptive optics integral field spectroscopy of the gravitational lens B1422+231},
  year      = {2014},
  issn      = {0035-8711},
  month     = jun,
  number    = {3},
  pages     = {2434--2445},
  volume    = {442},
  doi       = {10.1093/mnras/stu862},
  publisher = {Oxford University Press (OUP)},
}

@article{Birrer2021, doi = {10.21105/joss.03283}, url = {https://doi.org/10.21105/joss.03283}, year = {2021}, publisher = {The Open Journal}, volume = {6}, number = {62}, pages = {3283}, author = {Birrer, Simon and Shajib, Anowar J. and Gilman, Daniel and Galan, Aymeric and Aalbers, Jelle and Millon, Martin and Morgan, Robert and Pagano, Giulia and Park, Ji Won and Teodori, Luca and Tessore, Nicolas and Ueland, Madison and Van de Vyvere, Lyne and Wagner-Carena, Sebastian and Wempe, Ewoud and Yang, Lilan and Ding, Xuheng and Schmidt, Thomas and Sluse, Dominique and Zhang, Ming and Amara, Adam}, title = {lenstronomy II: A gravitational lensing software ecosystem}, journal = {Journal of Open Source Software} }

@Article{Schmidt2022,
  author    = {Schmidt, T and Treu, T and Birrer, S and Shajib, A J and Lemon, C and Millon, M and Sluse, D and Agnello, A and Anguita, T and Auger-Williams, M W and McMahon, R G and Motta, V and Schechter, P and Spiniello, C and Kayo, I and Courbin, F and Ertl, S and Fassnacht, C D and Frieman, J A and More, A and Schuldt, S and Suyu, S H and Aguena, M and Andrade-Oliveira, F and Annis, J and Bacon, D and Bertin, E and Brooks, D and Burke, D L and Carnero Rosell, A and Carrasco Kind, M and Carretero, J and Conselice, C and Costanzi, M and da Costa, L N and Pereira, M E S and De Vicente, J and Desai, S and Doel, P and Everett, S and Ferrero, I and Friedel, D and García-Bellido, J and Gaztanaga, E and Gruen, D and Gruendl, R A and Gschwend, J and Gutierrez, G and Hinton, S R and Hollowood, D L and Honscheid, K and James, D J and Kuehn, K and Lahav, O and Menanteau, F and Miquel, R and Palmese, A and Paz-Chinchón, F and Pieres, A and Plazas Malagón, A A and Prat, J and Rodriguez-Monroy, M and Romer, A K and Sanchez, E and Scarpine, V and Sevilla-Noarbe, I and Smith, M and Suchyta, E and Tarle, G and To, C and Varga, T N},
  journal   = {Monthly Notices of the Royal Astronomical Society},
  title     = {STRIDES: automated uniform models for 30 quadruply imaged quasars},
  year      = {2022},
  issn      = {1365-2966},
  month     = nov,
  number    = {1},
  pages     = {1260--1300},
  volume    = {518},
  doi       = {10.1093/mnras/stac2235},
  publisher = {Oxford University Press (OUP)},
}

@Article{Aghanim2020,
  author    = {Aghanim, N. and Akrami, Y. and Ashdown, M. and Aumont, J. and Baccigalupi, C. and Ballardini, M. and Banday, A. J. and Barreiro, R. B. and Bartolo, N. and Basak, S. and Battye, R. and Benabed, K. and Bernard, J.-P. and Bersanelli, M. and Bielewicz, P. and Bock, J. J. and Bond, J. R. and Borrill, J. and Bouchet, F. R. and Boulanger, F. and Bucher, M. and Burigana, C. and Butler, R. C. and Calabrese, E. and Cardoso, J.-F. and Carron, J. and Challinor, A. and Chiang, H. C. and Chluba, J. and Colombo, L. P. L. and Combet, C. and Contreras, D. and Crill, B. P. and Cuttaia, F. and de Bernardis, P. and de Zotti, G. and Delabrouille, J. and Delouis, J.-M. and Di Valentino, E. and Diego, J. M. and Doré, O. and Douspis, M. and Ducout, A. and Dupac, X. and Dusini, S. and Efstathiou, G. and Elsner, F. and Enßlin, T. A. and Eriksen, H. K. and Fantaye, Y. and Farhang, M. and Fergusson, J. and Fernandez-Cobos, R. and Finelli, F. and Forastieri, F. and Frailis, M. and Fraisse, A. A. and Franceschi, E. and Frolov, A. and Galeotta, S. and Galli, S. and Ganga, K. and Génova-Santos, R. T. and Gerbino, M. and Ghosh, T. and González-Nuevo, J. and Górski, K. M. and Gratton, S. and Gruppuso, A. and Gudmundsson, J. E. and Hamann, J. and Handley, W. and Hansen, F. K. and Herranz, D. and Hildebrandt, S. R. and Hivon, E. and Huang, Z. and Jaffe, A. H. and Jones, W. C. and Karakci, A. and Keihänen, E. and Keskitalo, R. and Kiiveri, K. and Kim, J. and Kisner, T. S. and Knox, L. and Krachmalnicoff, N. and Kunz, M. and Kurki-Suonio, H. and Lagache, G. and Lamarre, J.-M. and Lasenby, A. and Lattanzi, M. and Lawrence, C. R. and Le Jeune, M. and Lemos, P. and Lesgourgues, J. and Levrier, F. and Lewis, A. and Liguori, M. and Lilje, P. B. and Lilley, M. and Lindholm, V. and López-Caniego, M. and Lubin, P. M. and Ma, Y.-Z. and Macías-Pérez, J. F. and Maggio, G. and Maino, D. and Mandolesi, N. and Mangilli, A. and Marcos-Caballero, A. and Maris, M. and Martin, P. G. and Martinelli, M. and Martínez-González, E. and Matarrese, S. and Mauri, N. and McEwen, J. D. and Meinhold, P. R. and Melchiorri, A. and Mennella, A. and Migliaccio, M. and Millea, M. and Mitra, S. and Miville-Deschênes, M.-A. and Molinari, D. and Montier, L. and Morgante, G. and Moss, A. and Natoli, P. and Nørgaard-Nielsen, H. U. and Pagano, L. and Paoletti, D. and Partridge, B. and Patanchon, G. and Peiris, H. V. and Perrotta, F. and Pettorino, V. and Piacentini, F. and Polastri, L. and Polenta, G. and Puget, J.-L. and Rachen, J. P. and Reinecke, M. and Remazeilles, M. and Renzi, A. and Rocha, G. and Rosset, C. and Roudier, G. and Rubiño-Martín, J. A. and Ruiz-Granados, B. and Salvati, L. and Sandri, M. and Savelainen, M. and Scott, D. and Shellard, E. P. S. and Sirignano, C. and Sirri, G. and Spencer, L. D. and Sunyaev, R. and Suur-Uski, A.-S. and Tauber, J. A. and Tavagnacco, D. and Tenti, M. and Toffolatti, L. and Tomasi, M. and Trombetti, T. and Valenziano, L. and Valiviita, J. and Van Tent, B. and Vibert, L. and Vielva, P. and Villa, F. and Vittorio, N. and Wandelt, B. D. and Wehus, I. K. and White, M. and White, S. D. M. and Zacchei, A. and Zonca, A.},
  journal   = {Astronomy \& Astrophysics},
  title     = {Planck 2018 results: VI. Cosmological parameters},
  year      = {2020},
  issn      = {1432-0746},
  month     = sep,
  pages     = {A6},
  volume    = {641},
  doi       = {10.1051/0004-6361/201833910},
  publisher = {EDP Sciences},
}

@Article{Mao1998,
  author    = {Mao, Shude and Schneider, Peter},
  journal   = {Monthly Notices of the Royal Astronomical Society},
  title     = {Evidence for substructure in lens galaxies?},
  year      = {1998},
  issn      = {1365-2966},
  month     = apr,
  number    = {3},
  pages     = {587--594},
  volume    = {295},
  doi       = {10.1046/j.1365-8711.1998.01319.x},
  publisher = {Oxford University Press (OUP)},
}

@Article{Treu2010,
  author    = {Treu, Tommaso},
  journal   = {Annual Review of Astronomy and Astrophysics},
  title     = {Strong Lensing by Galaxies},
  year      = {2010},
  issn      = {1545-4282},
  month     = aug,
  number    = {1},
  pages     = {87--125},
  volume    = {48},
  doi       = {10.1146/annurev-astro-081309-130924},
  publisher = {Annual Reviews},
}

@Article{Nierenberg2024,
  author    = {Nierenberg, A M and Keeley, R E and Sluse, D and Gilman, D and Birrer, S and Treu, T and Abazajian, K N and Anguita, T and Benson, A J and Bennert, V N and Djorgovski, S G and Du, X and Fassnacht, C D and Hoenig, S F and Kusenko, A and Lemon, C and Malkan, M and Motta, V and Moustakas, L A and Stern, D and Wechsler, R H},
  journal   = {Monthly Notices of the Royal Astronomical Society},
  title     = {JWST lensed quasar dark matter survey – I. Description and first results},
  year      = {2024},
  issn      = {1365-2966},
  month     = feb,
  number    = {3},
  pages     = {2960--2971},
  volume    = {530},
  doi       = {10.1093/mnras/stae499},
  publisher = {Oxford University Press (OUP)},
}

@Article{Keeley2024,
  author    = {Keeley, Ryan E and Nierenberg, A M and Gilman, D and Gannon, C and Birrer, S and Treu, T and Benson, A J and Du, X and Abazajian, K N and Anguita, T and Bennert, V N and Djorgovski, S G and Gupta, K K and Hoenig, S F and Kusenko, A and Lemon, C and Malkan, M and Motta, V and Moustakas, L A and Oh, Maverick S H and Sluse, D and Stern, D and Wechsler, R H},
  journal   = {Monthly Notices of the Royal Astronomical Society},
  title     = {JWST lensed quasar dark matter survey – II. Strongest gravitational lensing limit on the dark matter free streaming length to date},
  year      = {2024},
  issn      = {1365-2966},
  month     = nov,
  number    = {2},
  pages     = {1652--1671},
  volume    = {535},
  doi       = {10.1093/mnras/stae2458},
  publisher = {Oxford University Press (OUP)},
}

@Misc{Keeley2025,
  author    = {Keeley, R. E. and Nierenberg, A. M. and Gilman, D. and Treu, T. and Du, X. and Gannon, C. and Mozumdar, P. and Wong, K. C. and Paugnat, H. and Birrer, S. and Malkan, M. and Benson, A. J. and Abazajian, K. N. and Anguita, T. and Bennert, V. N. and Djorgovski, S. G. and Hoenig, S. F. and Kusenko, A. and Larsson, H. R. and Morishita, T. and Motta, V. and Moustakas, L. A. and Sheu, W. and Sluse, D. and Stern, D. and Stiavelli, M. and Williams, D.},
  title     = {JWST Lensed Quasar Dark Matter Survey III: Dark Matter Sensitive Flux Ratios and Warm Dark Matter Constraint from the Full Sample},
  year      = {2025},
  copyright = {arXiv.org perpetual, non-exclusive license},
  doi       = {10.48550/ARXIV.2511.07765},
  publisher = {arXiv},
}

@Misc{Gilman2025,
  author    = {Gilman, D. and Nierenberg, A. M. and Treu, T. and Gannon, C. and Du, X. and Paugnat, H. and Birrer, S. and Benson, A. J. and Mozumdar, P. and Wong, K. C. and Williams, D. and Keeley, R. E. and Abazajian, K. N. and Anguita, T. and Bennert, V. N. and Djorgovski, S. G. and Hoenig, S. H. and Kusenko, A. and Malkan, M. and Morishita, T. and Motta, V. and Moustakas, L. A. and Sheu, W. and Sluse, D. and Stern, D. and Stiavelli, M.},
  title     = {JWST lensed quasar dark matter survey IV: Stringent warm dark matter constraints from the joint reconstruction of extended lensed arcs and quasar flux ratios},
  year      = {2025},
  copyright = {Creative Commons Attribution 4.0 International},
  doi       = {10.48550/ARXIV.2511.07513},
  publisher = {arXiv},
}

@Article{Dalal2002,
  author    = {Dalal, N. and Kochanek, C. S.},
  journal   = {The Astrophysical Journal},
  title     = {Direct Detection of Cold Dark Matter Substructure},
  year      = {2002},
  issn      = {1538-4357},
  month     = jun,
  number    = {1},
  pages     = {25--33},
  volume    = {572},
  doi       = {10.1086/340303},
  publisher = {American Astronomical Society},
}

@ARTICLE{Sersic1967,
       author = {{S{\'e}rsic}, J.~L.},
        title = "{Influence of the atmospheric and instrumental dispersion on the brightness distribution in a galaxy}",
      journal = {Boletin de la Asociacion Argentina de Astronomia La Plata Argentina},
         year = 1963,
        month = feb,
       volume = {6},
        pages = {41-43},
       adsurl = {https://ui.adsabs.harvard.edu/abs/1963BAAA....6...41S}
}

@Article{Schmitt2003,
  author    = {Schmitt, H. R. and Donley, J. L. and Antonucci, R. R. J. and Hutchings, J. B. and Kinney, A. L. and Pringle, J. E.},
  journal   = {The Astrophysical Journal},
  title     = {AHubble Space TelescopeSurvey of Extended [OIII] $\lambda$ 5007 A Emission in a Far‐Infrared–Selected Sample of Seyfert Galaxies: Results},
  year      = {2003},
  issn      = {1538-4357},
  month     = nov,
  number    = {2},
  pages     = {768--779},
  volume    = {597},
  doi       = {10.1086/381224},
  publisher = {American Astronomical Society},
}

@Article{Greene2011,
  author    = {Greene, Jenny E. and Zakamska, Nadia L. and Ho, Luis C. and Barth, Aaron J.},
  journal   = {The Astrophysical Journal},
  title     = {FEEDBACK IN LUMINOUS OBSCURED QUASARS},
  year      = {2011},
  issn      = {1538-4357},
  month     = apr,
  number    = {1},
  pages     = {9},
  volume    = {732},
  doi       = {10.1088/0004-637x/732/1/9},
  publisher = {American Astronomical Society},
}

@Article{Husemann2019,
  author    = {Husemann, Bernd and Bennert, Vardha N. and Jahnke, Knud and Davis, Timothy A. and Woo, Jong-Hak and Scharwächter, Julia and Schulze, Andreas and Gaspari, Massimo and Zwaan, Martin A.},
  journal   = {The Astrophysical Journal},
  title     = {Jet-driven Galaxy-scale Gas Outflows in the Hyperluminous Quasar 3C 273},
  year      = {2019},
  issn      = {1538-4357},
  month     = jul,
  number    = {2},
  pages     = {75},
  volume    = {879},
  doi       = {10.3847/1538-4357/ab24bc},
  publisher = {American Astronomical Society},
}

@Article{VillarMartin2011,
  author    = {Villar-Martín, M. and Humphrey, A. and Delgado, R. González and Colina, L. and Arribas, S.},
  journal   = {Monthly Notices of the Royal Astronomical Society},
  title     = {Ionized outflows in SDSS type 2 quasars at z $\sim$ 0.3-0.6: Ionized outflows in SDSS type 2 quasars},
  year      = {2011},
  issn      = {0035-8711},
  month     = sep,
  number    = {3},
  pages     = {2032--2042},
  volume    = {418},
  doi       = {10.1111/j.1365-2966.2011.19622.x},
  publisher = {Oxford University Press (OUP)},
}

@Article{MuellerSanchez2011,
  author    = {Müller-Sánchez, F. and Prieto, M. A. and Hicks, E. K. S. and Vives-Arias, H. and Davies, R. I. and Malkan, M. and Tacconi, L. J. and Genzel, R.},
  journal   = {The Astrophysical Journal},
  title     = {OUTFLOWS FROM ACTIVE GALACTIC NUCLEI: KINEMATICS OF THE NARROW-LINE AND CORONAL-LINE REGIONS IN SEYFERT GALAXIES,},
  year      = {2011},
  issn      = {1538-4357},
  month     = sep,
  number    = {2},
  pages     = {69},
  volume    = {739},
  doi       = {10.1088/0004-637x/739/2/69},
  publisher = {American Astronomical Society},
}

@Article{Husemann2014,
  author    = {Husemann, B. and Jahnke, K. and Sánchez, S. F. and Wisotzki, L. and Nugroho, D. and Kupko, D. and Schramm, M.},
  journal   = {Monthly Notices of the Royal Astronomical Society},
  title     = {Integral field spectroscopy of nearby QSOs – I. ENLR size–luminosity relation, ongoing star formation and resolved gas-phase metallicities},
  year      = {2014},
  issn      = {1365-2966},
  month     = jul,
  number    = {1},
  pages     = {755--783},
  volume    = {443},
  doi       = {10.1093/mnras/stu1167},
  publisher = {Oxford University Press (OUP)},
}

@Article{Humphrey2010,
  author    = {Humphrey, A. and Villar-Martín, M. and Sánchez, S. F. and Martínez-Sansigre, A. and Delgado, R. González and Pérez, E. and Tadhunter, C. and Pérez-Torres, M. A.},
  journal   = {Monthly Notices of the Royal Astronomical Society: Letters},
  title     = {Integral-field spectroscopy of type II QSOs at z = 0.3-0.4},
  year      = {2010},
  issn      = {1745-3925},
  month     = oct,
  number    = {1},
  pages     = {L1--L5},
  volume    = {408},
  doi       = {10.1111/j.1745-3933.2010.00906.x},
  publisher = {Oxford University Press (OUP)},
}

@Article{Boyarsky2009,
  author    = {Boyarsky, Alexey and Ruchayskiy, Oleg and Shaposhnikov, Mikhail},
  journal   = {Annual Review of Nuclear and Particle Science},
  title     = {The Role of Sterile Neutrinos in Cosmology and Astrophysics},
  year      = {2009},
  issn      = {1545-4134},
  month     = nov,
  number    = {1},
  pages     = {191--214},
  volume    = {59},
  doi       = {10.1146/annurev.nucl.010909.083654},
  publisher = {Annual Reviews},
}

@Article{Feng2010,
  author    = {Feng, Jonathan L.},
  journal   = {Annual Review of Astronomy and Astrophysics},
  title     = {Dark Matter Candidates from Particle Physics and Methods of Detection},
  year      = {2010},
  issn      = {1545-4282},
  month     = aug,
  number    = {1},
  pages     = {495--545},
  volume    = {48},
  doi       = {10.1146/annurev-astro-082708-101659},
  publisher = {Annual Reviews},
}

@Article{Lovell2014,
  author    = {Lovell, Mark R. and Frenk, Carlos S. and Eke, Vincent R. and Jenkins, Adrian and Gao, Liang and Theuns, Tom},
  journal   = {Monthly Notices of the Royal Astronomical Society},
  title     = {The properties of warm dark matter haloes},
  year      = {2014},
  issn      = {0035-8711},
  month     = feb,
  number    = {1},
  pages     = {300--317},
  volume    = {439},
  doi       = {10.1093/mnras/stt2431},
  publisher = {Oxford University Press (OUP)},
}

@Article{Weinberg2015,
  author    = {Weinberg, David H. and Bullock, James S. and Governato, Fabio and Kuzio de Naray, Rachel and Peter, Annika H. G.},
  journal   = {Proceedings of the National Academy of Sciences},
  title     = {Cold dark matter: Controversies on small scales},
  year      = {2015},
  issn      = {1091-6490},
  month     = feb,
  number    = {40},
  pages     = {12249--12255},
  volume    = {112},
  doi       = {10.1073/pnas.1308716112},
  publisher = {Proceedings of the National Academy of Sciences},
}

\end{document}